\documentclass[amsmath,amssymb,reprint,aps,prd,eqsecnum,unsortedaddress]{revtex4-2}
 
\usepackage{graphicx}
\usepackage{dcolumn}
\usepackage{bm}
\usepackage{mathrsfs}
\usepackage{hyperref}
\usepackage[dvipsnames]{xcolor}
\usepackage{todonotes}
\usepackage{comment}

\newcommand{\subE}{\textrm{\tiny{E}}}

\newcommand{\supR}{\textrm{\tiny{R}}}
\newcommand{\supP}{\textrm{\tiny{P}}}
\newcommand{\supL}{\textrm{\tiny{L}}}
\newcommand{\subF}{\textrm{\tiny{F}}}
\newcommand{\sy}{\textrm{\tiny{(S)}}}
\newcommand{\asy}{\textrm{\tiny{(A)}}}

\allowdisplaybreaks

\begin{document}

\title{Mode-sum prescription for renormalized expectation values\\ for a charged quantum scalar field on a charged black hole}

\author{Cormac Breen}

\email{cormac.breen@tudublin.ie}

\affiliation{School of Mathematics and Statistics,
Technological University Dublin,
Grangegorman,
Dublin 7, Ireland}

\author{George Montagnon}

\email{GJMontagnon1@sheffield.ac.uk}

\affiliation{School of Mathematical and Physical Sciences,
The University of Sheffield,
Hicks Building,
Hounsfield Road,
Sheffield. S3 7RH United Kingdom}

\author{Peter Taylor}

\email{peter.taylor@dcu.ie}

\affiliation{Centre for Astrophysics and Relativity,
School of Mathematical Sciences,
Dublin City University,
Glasnevin,
Dublin 9, Ireland.}

\author{Elizabeth Winstanley}

\email{E.Winstanley@sheffield.ac.uk}

\affiliation{School of Mathematical and Physical Sciences,
The University of Sheffield,
Hicks Building,
Hounsfield Road,
Sheffield. S3 7RH United Kingdom}

\date{\today}

\begin{abstract}
We present a new mode-sum prescription for the efficient computation of renormalized expectation values for a massive, charged, quantum scalar field propagating on a curved space-time background.
Our method is applicable to any static, spherically-symmetric, four-dimensional space-time with a time-independent, background electrostatic potential and can be used to find the renormalized scalar condensate, current and stress-energy tensor.
As an explicit example, we present a calculation of these quantities for a charged scalar field in the Hartle-Hawking state on a Reissner-Nordstr\"om black hole background.
\end{abstract}

\maketitle

\section{Introduction}
\label{sec:intro}

Hawking's discovery \cite{Hawking:1975vcx} that black holes emit quantum thermal radiation provoked many deep questions about the interplay between general relativity and quantum field theory on black hole space-times.   
Taking a semiclassical approach, the background geometry is a solution of the classical Einstein equations, possibly with additional background matter sources having (classical) stress-energy tensor $T_{\alpha \beta }^{{\rm {(C)}}}$:
\begin{equation}
    G_{\alpha \beta } + \Lambda g_{\alpha \beta }  = T_{\alpha \beta }^{{\rm {(C)}}} ,
\end{equation}
where $G_{\alpha \beta }$ is the (classical) Einstein tensor of the background metric, $\Lambda $ the cosmological constant and $g_{\alpha \beta }$ the metric (here and throughout this paper, we use units in which $8\pi G = c = \hbar = k_{\rm {B}}=1$).
Having fixed the space-time and matter fields, one can then study the behaviour of a quantum field on this fixed background.

However, such a framework is a lowest-order approximation. 
All matter and energy gravitates, including a quantum field on a curved space-time background.
This back-reaction process results in the evaporation of a black hole due to Hawking radiation \cite{Hawking:1975vcx}.
The back-reaction of the quantum field on the space-time geometry is described by the semiclassical Einstein equations
\begin{equation}
    G_{\alpha \beta } + \Lambda g_{\alpha \beta }  = T_{\alpha \beta }^{{\rm {(C)}}}  + \langle {\hat {T}}_{\alpha \beta } \rangle ,
\end{equation}
where $\langle {\hat {T}}_{\alpha \beta } \rangle $ is the expectation value of the stress-energy tensor of the quantum field.
Finding $\langle {\hat {T}}_{\alpha  \beta } \rangle $ is therefore of central importance; it encodes detailed properties of the quantum field itself as well as acting as a source term for modifications of the underlying space-time.

In this paper we focus on the situation where the classical background matter is an electromagnetic field whose Faraday tensor $F^{\alpha \beta }=\nabla ^{\alpha }A^{\beta }-\nabla ^{\beta }A^{\alpha }$ (where $A^{\alpha }$ is the electromagnetic potential) satisfies Maxwell's equations
\begin{equation}
    \nabla _{\beta }F^{\beta \alpha } = 0,
\end{equation}
and the quantum field possesses a nonzero charge, so that it interacts with the electromagnetic field as well as the background geometry.
The charge of the quantum field introduces an effective chemical potential into the Hawking radiation \cite{Gibbons:1975kk,Hawking:1975vcx}, thereby modifying the emission and extracting charge as well as mass from a black hole.
A charged quantum scalar field will have an effect, not only on the background space-time, but also on the background electromagnetic field, since it will provide a source term in Maxwell's equations:
\begin{equation}
    \nabla _{\beta }F^{\beta \alpha } = 4\pi \langle {\hat {J}}^{\alpha } \rangle ,
\end{equation}
where $\langle {\hat {J}}^{\alpha }\rangle $ is the expectation value of the current operator for the quantum field (and we are using Gaussian units).  
Therefore, to study the effect of a charged quantum field, computation of $\langle {\hat {J}}^{\alpha } \rangle$ as well as $\langle {\hat {T}}_{\alpha \beta } \rangle $ is required.

Computation of either $\langle {\hat {J}}^{\alpha } \rangle$ or $\langle {\hat {T}}_{\alpha \beta } \rangle $ is extremely challenging; both involve the products of field operators at the same space-time point and are therefore formally divergent, resulting in the need to employ a regularization and renormalization prescription.
The divergences can be regularized by point-splitting \cite{Christensen:1976vb,Christensen:1978yd}, as the Green function describing the quantum field depends on two space-time points and is regular when those points are separated. 
Formally, the two expectation values can be found by applying a suitable differential operator (see, for example, \cite{Decanini:2005eg,Balakumar:2019djw}) to the Green function, and then taking the coincidence limit in which the space-time points are brought together.
The divergences in the expectation values arise from divergences in the Green function in the coincidence limit.
For physical quantum states, the singularities in the Green function can be renormalized by subtracting a distribution given by the Hadamard parametrix  \cite{Fulling:1978ht} from the Green function, before applying any differential operator and bringing the space-time points together.

In practice, this procedure is technically difficult: the Green function is typically given as an infinite sum over mode solutions of the classical field equation, each of which can only be computed numerically, while the Hadamard parametrix is a geometric quantity involving coefficients which have Taylor series expansions in the coordinate separation of the points. 
The first methodology for a practical computation of the renormalized expectation value of the stress-energy tensor $\langle {\hat {T}}_{\alpha \beta } \rangle _{\rm {ren}}$ (RSET) for a neutral quantum scalar field was developed by Howard and Candelas \cite{Howard:1984qp,Howard:1984ttx} 
for a Schwarzschild black hole space-time. 
This was subsequently refined in \cite{Anderson:1993if,Anderson:1994hg,Breen:2011aa} to give a method valid for a neutral scalar field on any static, spherically symmetric background geometry.
The key feature of this approach is to work on the Euclideanized version of the space-time, performing a Wick rotation of the time-coordinate. 
This has the advantage that the Euclidean Green function is singular only in the coincidence limit, 
unlike the Green function on the Lorentzian space-time, which requires an $i\epsilon $ prescription in order to avoid singularities when the separated points are connected by a null geodesic.
Unfortunately, the method of \cite{Howard:1984qp,Howard:1984ttx,Anderson:1993if,Anderson:1994hg,Breen:2011aa} 
is cumbersome to implement except in the simplest cases as it relies on a nonuniform WKB approximation for the scalar field modes. 

The development, over the past ten years, of two new methods for computing renormalized expectation values has greatly facilitated these calculations and enabled the properties of a neutral quantum scalar field on a much wider range of black hole space-times to be explored.
The ``pragmatic mode-sum'' method \cite{Levi:2016quh,Levi:2016paz} works on the original Lorentzian space-time and is notable for its applicability to rotating as well as static black holes \cite{Levi:2016exv}.
In contrast, the ``extended coordinates'' method \cite{Taylor:2022sly} involves the Euclidean space-time.
The key feature of the latter approach is that the Hadamard parametrix to any order in the coordinate separation can be expressed as a mode-sum, resulting in a renormalized Green function that converges to any order we desire; the rate of convergence being directly linked to the order of the parametrix we subtract.
This method also has the advantage that it is approximately uniform, including close to the horizon.

Progress in the computation of renormalized expectation values for a charged scalar field has been more limited.
On an electrically-charged Reissner-Nordstr\"om-de Sitter black hole, the renormormalized current $\langle {\hat {J}}^{\alpha }\rangle _{\rm {ren}}$ was computed in \cite{Klein:2021les,Klein:2021ctt}.
Working on the Lorentzian space-time, making a suitable choice of gauge at each space-time point, and a suitable choice of point-splitting, it is shown that only finite (or zero) renormalization terms are required for the current, which facilitates the computation.
In \cite{Klein:2021ctt} one component of the RSET $\langle {\hat {T}}_{\alpha \beta }\rangle _{\rm {ren}}$ is also computed, which similarly requires only finite or vanishing renormalization terms. 
Other calculations in the literature have been performed on an electrically-charged Reissner-Nordstr\"om black hole, and only involve quantities which do not require renormalization, such as
the components $\langle {\hat {J}}^{r}\rangle $ and $\langle {\hat {T}}_{tr} \rangle $ \cite{Balakumar:2020gli} and  differences in expectation values between two quantum states \cite{Balakumar:2022yvx}.

To explore in detail the behaviour of a charged quantum scalar field on a charged black hole background, including the back-reaction, a methodology for the computation of all components of the renormalized current and RSET is required. 
First steps in this direction were taken many years ago \cite{Herman:1995hm,Herman:1998dz} but until now have not been developed into a practical implementation.
Our purpose in this paper is to present a new methodology for the computation of all components of both the current and RSET, building on the ``extended coordinates'' approach of \cite{Taylor:2016edd,Taylor:2017sux,Taylor:2022sly}.

Our method is applicable to any static, spherically symmetric geometry with a background electrostatic potential, and a quantum scalar field with arbitrary mass, charge and coupling to the Ricci scalar curvature. 
As in \cite{Taylor:2016edd,Taylor:2017sux,Taylor:2022sly}, we work on Euclidean space-time, 
and first construct the Euclidean Green function for a quantum charged scalar field, in Sec.~\ref{sec:Green}.
The Hadamard parametrix for this set-up has already been derived in \cite{Balakumar:2019djw}, and in Sec.~\ref{sec:Hadamard} we find this parametrix in terms of ``extended coordinates'' as defined in \cite{Taylor:2016edd,Taylor:2017sux,Taylor:2022sly}.
In Sec.~\ref{sec:expectation} we write the Hadamard parametrix as a mode sum and use this to perform the renormalization mode-by-mode, obtaining expressions for the renormalized scalar condensate (the square of the scalar field), current, and RSET. 
As a demonstration of the efficacy of this method, in Sec.~\ref{sec:RNHH} we perform a computation of these quantities for a massive charged scalar field in the Hartle-Hawking state \cite{Hartle:1976tp} on a Reissner-Nordstr\"om black hole.  
We close, in Sec.~\ref{sec:conc}, with our conclusions.

\section{Euclidean Green function}
\label{sec:Green}

The ``extended coordinate" approach to implementing Hadamard renormalization \cite{Taylor:2016edd,Taylor:2017sux,Taylor:2022sly} is applicable to static, spherically symmetric space-times with well-defined Euclidean sections. 
Such space-times are described by line elements of the form
\begin{equation}
    ds^{2}=f(r) \, d\tau^{2}+\frac{1}{f(r)}dr^{2}+r^{2}d\Omega^{2},
    \label{eq:metric}
\end{equation}
derived via a Wick rotation ($t\to-i\tau$) of the original Lorentzian space-time. 
Here, $f(r)$ denotes some metric function which, for the time being we leave unspecified, and $d\Omega^{2}$ is the usual line element of the two-sphere. 
When (\ref{eq:metric}) describes a black hole space-time,
in order to describe quantum states that are regular on the event horizon, it is necessary to impose a periodicity in the $\tau$ coordinate, $\tau=\tau+2\pi/\kappa$ with $\kappa$ the event horizon surface gravity, in order to avoid the existence of a conical singularity. 

In this paper we are concerned with calculating the renormalized expectation values of observables for a charged scalar field. 
The charged scalar field satisfies the equation of motion
\begin{equation}
\label{eq:waveeqn}
    \left[ D_{\alpha }D^{\alpha  } - \mu ^{2} -\xi R \right] \Phi = 0,
\end{equation}
where $\mu$ is the field mass, $R$ is the Ricci scalar, $\xi$ is the coupling to the space-time curvature, $D_{\alpha }=\nabla _{\alpha }-iqA_{\alpha }$  is the gauge covariant derivative, $q$ is the scalar field charge, and $A$ is the gauge field which we assume is of the form $A=A_{t}(r)\, dt$, with $A_{t}(r)$ a real function of $r$. 
On the Euclidean section (\ref{eq:waveeqn}) takes the form \cite{Herman:1998dz}
\begin{equation}
    \left[ \square_{\subE}-2iqA^{\tau}\nabla_{\tau}-q^{2}A^{\tau}A_{\tau}-\mu^{2}-\xi R\right] \Phi=0,
    \label{eq:waveEucl}
\end{equation}
where $\square_{\subE}$ is the d'Alembertian operator, emphasising that this differential equation is now of elliptic type, and $A_{\tau}$ is given, through Wick rotation, by $A_{\tau}=-iA_{t}$ and hence is purely imaginary \cite{Braden:1990hw,Herman:1998dz}. 
As a result, the differential operator on the left-hand-side of (\ref{eq:waveEucl}) is real, unlike that on the left-hand-side of the equation of motion (\ref{eq:waveeqn}) on the Lorentzian space-time.
From henceforth, we will omit the subscript E from covariant derivatives, which should be assumed to be taken with respect to the Euclidean metric (\ref{eq:metric}) unless otherwise stated. 

In black hole space-time applications, this framework most naturally applies to a field considered in the Hartle-Hawking state \cite{Hartle:1976tp}, thus we consider a Euclidean Green function  $G_{\subE}(x,x')$ corresponding to a thermal state at temperature $T=\kappa/2\pi$. 
This may be expanded as 
\begin{align}
\label{eq:GE}
    &G_{\subE}(x,x')\nonumber \\ 
    &=\frac{1}{8\pi^{2}}\sum_{l=0}^{\infty}(2l+1)P_{l}(\cos\gamma)\sum_{n=-\infty}^{\infty}e^{in\kappa\Delta\tau}g_{nl}(r,r'),
\end{align}
where
$\Delta \tau = \tau - \tau '$, the angular separation $\gamma $ is given by 
\begin{equation}
    \cos \gamma = \cos \theta \cos \theta ' + \sin \theta \sin \theta ' \cos \left( \phi - \phi ' \right) ,
\end{equation}
$P_{l}(\cos \gamma )$ is a Legendre polynomial,
the radial Green function $g_{nl}(r,r')$ is
\begin{equation}
\label{eq:gnl}
    g_{nl}(r,r')=\kappa\mathcal{N}_{nl}p_{nl}(r_{<})q_{nl}(r_{>}),
\end{equation}
with $r_{<}=\min\{r,r'\}$, $r_{>}=\max\{r,r'\}$ and $\mathcal{N}_{nl}$ is the normalization constant
\begin{align}
    \mathcal{N}_{nl}=-\frac{1}{r^{2}f(r)\mathcal{W}\{ p_{nl}(r), q_{nl}(r)\} },
\end{align}
with $\mathcal{W}\{.,.\}$ denoting the Wronskian. 
The functions $p_{nl}(r)$ and $q_{nl}(r)$ are solutions of the homogeneous radial equation
\begin{align}
\label{eq:Rad}
     \Bigg[\frac{d}{dr}\left(r^{2}f(r)\frac{d}{dr}\right)&-\frac{r^{2}}{f(r)}\left(n\kappa-qA_{\tau}\right)^{2}
    \nonumber \\ 
    -r^{2}\left(\mu^{2}+\xi R\right)&-l(l+1)\Bigg]Y_{nl}(r)=0,
\end{align}
satisfying appropriate boundary conditions. 
We note that the differential operator on the left-hand-side of (\ref{eq:Rad}) is complex due to the presence of $A_{\tau }$ which is imaginary, however the Euclidean Green function $G_{\subE}(x,x')$ is real.
The solutions $Y_{nl}(r)$ of (\ref{eq:Rad}) with positive and negative $n$ are related by
\begin{equation}
Y_{-nl}(r) = Y_{nl}^{*}(r),
    \label{eq:npm}
\end{equation}
where ${}^{*}$ denotes the complex conjugate. 
Solutions $Y_{0l}(r)$ are thus real.

For our later analysis, it is convenient to decompose the Euclidean Green function into its symmetric and antisymmetric parts, namely
\begin{subequations}
\label{eq:Gparts}
\begin{align}
    G_{\subE}(x,x')=G^{\sy}(x,x')+G^{\asy}(x,x'),
\end{align}
where
\begin{align}
\label{eq:GESym}
    G^{\sy}(x,x')=&\frac{1}{8\pi^{2}}\sum_{l=0}^{\infty}(2l+1)P_{l}(\cos\gamma)\nonumber\\
    &\quad\times\sum_{n=-\infty}^{\infty}\cos(n\kappa\Delta\tau)g_{nl}(r,r'),\\
    \label{eq:GEASym}
     G^{\asy}(x,x')=&\frac{i}{8\pi^{2}}\sum_{l=0}^{\infty}(2l+1)P_{l}(\cos\gamma)\nonumber\\
    &\quad\times\sum_{n=-\infty}^{\infty}\sin(n\kappa\Delta\tau)g_{nl}(r,r').
\end{align}
\end{subequations}
Although $g_{nl}(r,r')$ (\ref{eq:gnl}) is complex, both the symmetric and antisymmetric parts of the Euclidean Green function are real, since (\ref{eq:npm}) implies that
\begin{equation}
\label{eq:gnpm}
    g_{-nl}(r,r') = g_{nl}^{*}(r,r').
\end{equation}

Our desired expectation values can be derived from coincidence limits involving the symmetric and antisymmetric parts of the Euclidean Green function (\ref{eq:Gparts}).
However, at present, na\"{i}ve coincidence limits of (\ref{eq:GESym}) and the time derivative of (\ref{eq:GEASym}) (which is necessary for the computation of the current) lead to divergences in the sense that the mode sums above do not converge in this limit.
In the following sections we show how the ``extended coordinate" method \cite{Taylor:2016edd,Taylor:2017sux,Taylor:2022sly} can be utilised to remove these divergences and derive sensible results for the expectation values of observables of the charged scalar field. 

\section{Expansion of the Hadamard Parametrix}
\label{sec:Hadamard}

In this section, we will describe the renormalization prescription employed to extract meaningful finite results from the na\"{i}ve divergent expectation values of quantities that involve the Euclidean Green function (\ref{eq:Gparts}). 
The quantities of interest in this paper are the charged scalar condensate, the expectation value of the current and the expectation value of the stress-energy tensor, all of which require the coincidence limit of the Euclidean Green function and its derivatives. 
From the previous section, we note that the mode-sums in the representation of the Euclidean Green function (\ref{eq:GESym}) (and the time derivative of (\ref{eq:GEASym})) do not converge at coincidence and we require a meaningful way to render these mode-sums finite.

At least formally, the prescription for subtracting these divergences is well understood \cite{Decanini:2005eg,Balakumar:2019djw}. 
Noting that the short-distance singularity structure of the Euclidean Green function is universal (independent of the quantum state) provided the quantum state satisfies the Hadamard property, the divergences are encoded in the Hadamard parametrix, which is a locally constructed bidistribution such that the difference between the Green function and the Hadamard parametrix satisfies an inhomogeneous wave equation with a regular source term. 

On the Euclidean section of a four-dimensional space-time, the Hadamard parametrix has the following universal form \cite{Decanini:2005eg,Balakumar:2019djw} 
\begin{align}
\label{eq:KHad}
    K(x,x')=
    \frac{1}{8\pi^{2}}\left[\frac{U(x,x')}{\sigma(x,x')}+V(x,x')\log{\left(\frac{2\sigma(x,x')}{L^{2}}\right)}\right],
\end{align}
where $\sigma(x,x')$ is Synge's world function corresponding to half the square of the geodesic distance between the points $x$ and $x'$ (which is positive definite on a Euclidean section when $x\ne x'$). 
The biscalars $U(x,x')$ and $V(x,x')$ are regular in the coincidence limit and are constructed locally from the metric and its derivatives \cite{Balakumar:2019djw}. 
An arbitrary length scale $L$ has been inserted into the $\log $ term  in (\ref{eq:KHad}) to make the argument dimensionless. 
This arbitrariness is part of the well-known renormalization ambiguity \cite{Decanini:2005eg,Balakumar:2019djw} and is a manifestation of the fact that the biscalar $V(x,x')$ is a solution to the charged scalar field homogeneous wave equation \cite{Balakumar:2019djw}
\begin{align}
\label{eq:Veqn}
    \left[ D_{\alpha }D^{\alpha }-(\mu^{2}+\xi R)\right] &V(x,x')=0,
\end{align}
and so we are free to add multiples of $V(x,x')$ to any parametrix. 
The biscalar $U(x,x')$ satisfies the transport equation  \cite{Balakumar:2019djw} 
\begin{align}
\label{eq:Ueqn}
\left[ 2\sigma^{\alpha }D_{\alpha }+\Box\sigma-4\right] & U(x,x')=0,
\end{align}
subject to the boundary condition
\begin{equation}
    U(x,x)=1.
\end{equation}
Here and throughout, we have adopted the common notation $\sigma_{\mu}:=\nabla_{\mu}\sigma$. 

For a neutral scalar field, the two-point function is symmetric in the arguments $x$ and $x'$ and hence $U(x,x')$ and $V(x,x')$ are also symmetric in that case. 
However, for a charged scalar, this is no longer the case, as can be seen in (\ref{eq:Gparts}).
While the reduced symmetry renders (\ref{eq:Ueqn}) more difficult to solve in practice than the neutral case, it is still a transport equation along the geodesic connecting $x$ and $x'$ and readily solved as a covariant Taylor series about one of the points. 

In the same way, we can expand $V(x,x')$ as
\begin{equation}
\label{eq:Vdef}
    V(x,x') = \sum_{k=0}^{\infty}V_{k}(x,x')\sigma^{k},
\end{equation}
in the charged scalar field equation (\ref{eq:Veqn}) and equating equal powers in $\sigma$ gives a set of transport equations for each $V_{k+1}(x,x')$ (with $k\ge 0$), namely \cite{Balakumar:2019djw}, 
\begin{subequations}
  \label{eq:Vkeqn}  
\begin{multline}
(k+1)\left[2\sigma^{\alpha }D_{\alpha }+\Box\sigma+2k\right]V_{k+1}(x,x') \\
+ \left[ D_{\alpha }D^{\alpha }-(\mu^{2}+\xi R)\right] V_{k}(x,x') =0,
\label{eq:Vkeqngen}
\end{multline}
together with the transport equation for $V_{0}$
\begin{multline}
[ 2\sigma^{\alpha }D_{\alpha }+\Box\sigma-2]V_{0}(x,x') \\
   +[ D_{\alpha }D^{\alpha }-(\mu^{2}+\xi\,R)] U(x,x')=0.
   \label{eq:V0eqn}
\end{multline}
\end{subequations}
The $V_{k}(x,x')$ are now easily found as covariant Taylor series, subject to boundary conditions obtained by taking the coincidence limits of the respective equations.

The ``extended coordinate'' approach to renormalization involves expanding the biscalars $\sigma (x,x')$, $U(x,x')$ and $V_{k}(x,x')$ in terms of a judiciously chosen set of ``extended coordinates" $\varpi$ and $s$:
\begin{subequations}
\label{eq:extexp}
\begin{align}
    \sigma(x,x') &= \sum_{a,b,c}\sigma_{abc}(r)\varpi^{a}\Delta r^{b}s^{c},
    \label{eq:sigmaexp}
    \\
    U(x,x') &= \sum_{a,b,c}u_{abc}(r)\varpi^{a}\Delta r^{b}s^{c},
    \label{eq:Uexp}\\
    V_{k}(x,x') &= \sum_{a,b,c}v^{(k)}_{abc}(r)\varpi^{a}\Delta r^{b}s^{c},
    \label{eq:Vexp}
\end{align}
\end{subequations}
where
\begin{subequations}
\label{eq:extendedcoords}
\begin{align}
    \varpi&=\frac{2}{\kappa}\sin\left(\frac{\kappa\Delta\tau}{2}\right), 
    \label{eq:varpi}
    \\ s^{2}&=f(r)\varpi^{2}+2r^{2}(1-\cos\gamma),
\end{align}
\end{subequations}
and $\Delta r=r-r'$ is the separation of the points in the radial direction.
The coefficients in the expansions (\ref{eq:extexp}) can then be found by substituting (\ref{eq:extexp}) into the relevant equations (\ref{eq:Ueqn}, \ref{eq:Vkeqn}) and solving order by order, treating $\varpi\sim s\sim\Delta r\sim\mathcal{O}(\epsilon)$. 
We begin by substituting the ansatz (\ref{eq:sigmaexp}) into the defining relation
\begin{equation}
    2\sigma = \sigma _{\alpha }\sigma ^{\alpha },
\end{equation}
writing the derivatives of the extended coordinates (\ref{eq:extendedcoords}) as series in $(\varpi, s, \Delta r)$. 
Solving order by order in $\epsilon$ gives the coefficients $\sigma _{abc}(r)$, which are unchanged from those arising in the neutral case \cite{Taylor:2016edd,Taylor:2017sux,Taylor:2022sly}.
This expansion for $\sigma (x,x')$ is then used, together with (\ref{eq:Ueqn}), to find the coefficients in $U(x,x')$ (\ref{eq:Uexp}). 
Next, having the expansions of $\sigma (x,x')$ and $U(x,x')$, we use these in (\ref{eq:V0eqn}) to find the coefficients in the expansion of $V_{0}(x,x')$, and then, having $V_{k}(x,x')$, we can use (\ref{eq:Vkeqngen}) to find the coefficients in the expansion of $V_{k+1}(x,x')$.
Finally, the resulting expansions (\ref{eq:extexp}) are combined into the terms appearing in (\ref{eq:KHad}).  
The algebraic manipulations required to find the expansions (\ref{eq:extexp}) to the order required are extremely lengthy, and are therefore performed in {\tt {Mathematica}}.
Further details can be found in \cite{Taylor:2016edd,Taylor:2017sux,Taylor:2022sly}.

A key feature of this particular choice of ``extended coordinates'' is that in the limit $\Delta r\to 0$, we have $s^{2}=2\sigma+\mathcal{O}(\epsilon^{3})$. 
Note also that, although the Hadamard parametrix itself (\ref{eq:KHad}) is independent of the quantum state, this particular choice of expansion parameters (\ref{eq:extendedcoords}) is adapted specifically to quantum fields in thermal states. 
This is easily seen in the fact that the extended time coordinate $\varpi$ (\ref{eq:varpi}) is periodic with periodicity related to $\kappa $. 
This is an essential feature of the method, since the goal is to obtain a mode-sum representation of the parametrix (\ref{eq:KHad}) that has the same form as the Euclidean Green function (\ref{eq:GE}) so that a mode-by-mode subtraction can be performed.

For a neutral scalar field, the symmetry of the biscalars implies that only even powers of $\varpi$ and $s$ appear in the expansion of the Hadamard parametrix. 
For charged scalar fields, 
while the parametrix contains only even powers of $s$, odd powers of $\varpi$ also arise. 

Let us deal first with the direct part of the parametrix, $U(x,x')/\sigma(x,x')$. 
If we expand the direct part in our extended coordinates up to order $\mathcal{O}(\epsilon^{2m})$ and then take the partial coincidence limit $\Delta r=0$, we obtain
\begin{align}
    \frac{U(x,x')}{\sigma(x,x')}=&\sum_{a=0}^{m}\sum_{b=0}^{a}\mathcal{DE}_{ab}^{(\supR)}(r)\frac{\varpi^{2a+2b}}{s^{2b+2}}\nonumber\\
 &+\sum_{a=0}^{m-1}\sum_{b=0}^{a}\mathcal{DO}_{ab}^{(\supR)}(r)\frac{\varpi^{2a+2b+1}}{s^{2b+2}}\nonumber\\
&+\sum_{a=1}^{m}\sum_{b=1}^{a}\mathcal{DE}_{ab}^{(\supP)}(r)\varpi^{2a-2b}s^{2b-2}\nonumber\\
&+\sum_{a=1}^{m-1}\sum_{b=1}^{a}\mathcal{DO}_{ab}^{(\supP)}(r)\varpi^{2a-2b+1}s^{2b-2}
\nonumber \\ & \qquad +\mathcal{O}(\epsilon^{2m}).
\end{align}
The functions $\mathcal{DE}^{(\supR)}_{ab}(r)$ correspond to the coefficients of terms appearing in the direct part of the Hadamard parametrix that are rational in $\varpi$ and $s^{2}$ and even in $\varpi$, while $\mathcal{DO}^{(\supP)}_{ab}(r)$ are the coefficients of terms that are polynomial in $\varpi$ and $s^{2}$ and odd in $\varpi$, etc. 
So the labelling convention is: $\mathcal{D}$ refers to the direct part, $\mathcal{O}$/$\mathcal{E}$ refers to odd/even in $\varpi$, the superscript $(\textrm{R})$/$(\textrm{P})$ refers to whether the term is rational/polynomial in $\varpi$ and $s^{2}$. In App.~\ref{app:hadamardcoeff}, we list the first few of these coefficients, higher order coefficients can be found in the attached {\tt {Mathematica}} notebook.

We can also similarly expand the so-called tail part of the Hadamard parametrix $V(x,x)\log (2\sigma/L^{2})$, which after again taking the radial points at coincidence gives
\begin{align} 
V(x,x')&\log\left(\frac{2\sigma(x,x')}{L^{2}}\right)
\nonumber \\ 
 = &\sum_{a=1}^{m-1}\sum_{b=0}^{a-1}\mathcal{TE}_{ab}^{(\supR)}(r)\frac{\varpi^{2a+2b+2}}{s^{2b+2}}\nonumber\\
 &+\sum_{a=1}^{m-2}\sum_{b=0}^{a-1}\mathcal{TO}_{ab}^{(\supR)}(r)\frac{\varpi^{2a+2b+3}}{s^{2b+2}}\nonumber\\
&+\sum_{a=0}^{m-1}\sum_{b=0}^{a}\mathcal{TE}_{ab}^{(\supL)}(r)\varpi^{2b}s^{2a-2b}\log(s^{2}/L^{2})\nonumber\\
&+\sum_{a=0}^{m-2}\sum_{b=0}^{a}\mathcal{TO}_{ab}^{(\supL)}(r)\varpi^{2b+1}s^{2a-2b}\log(s^{2}/L^{2})\nonumber\\
&+\sum_{a=1}^{m-1}\sum_{b=0}^{a}\mathcal{TE}_{ab}^{(\supP)}(r)\varpi^{2b}s^{2a-2b}\nonumber\\
&+\sum_{a=1}^{m-2}\sum_{b=0}^{a}\mathcal{TO}_{ab}^{(\supP)}(r)\varpi^{2b+1}s^{2a-2b}
\nonumber \\ & \qquad  +\mathcal{O}(\epsilon^{2m}\log\epsilon).
\end{align}
The coefficients are now labelled by a $\mathcal{T}$ to indicate these come from the tail part of the parametrix and the superscript $(\mathrm{L})$ indicates coefficients of terms that contain a logarithm. 
Again, the first few of these coefficients are listed in App.~\ref{app:hadamardcoeff}.

It will be convenient for later purposes to decompose the Hadamard parametrix (\ref{eq:KHad}) as
\begin{subequations}
\label{eq:KHadext}
\begin{align}
    K(x,x')=K^{\sy}(x,x')+K^{\asy}(x,x')
\end{align}
where $K^{\sy}(x,x')$ and $K^{\asy}(x,x')$ are the symmetric and antisymmetric parts, respectively. Explicitly, the expansions of these in the radial coincidence limit are
\begin{align}
\label{eq:HadamardSym}
K^{\sy}(x,x')= & \frac{1}{8\pi^{2}}\Bigg\{\sum_{a=0}^{m}\sum_{b=0}^{a}\mathcal{DE}_{ab}^{(\supR)}(r)\frac{\varpi^{2a+2b}}{s^{2b+2}}\nonumber\\
&+\sum_{a=1}^{m}\sum_{b=1}^{a}\mathcal{DE}_{ab}^{(\supP)}(r)\varpi^{2a-2b}s^{2b-2}\nonumber\\
&+\sum_{a=1}^{m-1}\sum_{b=0}^{a-1}\mathcal{TE}_{ab}^{(\supR)}(r)\frac{\varpi^{2a+2b+2}}{s^{2b+2}}\nonumber\\
&+\sum_{a=0}^{m-1}\sum_{b=0}^{a}\mathcal{TE}_{ab}^{(\supL)}(r)\varpi^{2b}s^{2a-2b}\log(s^{2}/L^{2})\nonumber\\
&+\sum_{a=1}^{m-1}\sum_{b=0}^{a}\mathcal{TE}_{ab}^{(\supP)}(r)\varpi^{2b}s^{2a-2b}\Bigg\}
\nonumber \\ &  \qquad +\mathcal{O}(\epsilon^{2m}\log\epsilon),\\
\label{eq:HadamardASym}
K^{\asy}(x,x')= &\frac{1}{8\pi^{2}}\Bigg\{\sum_{a=0}^{m-1}\sum_{b=0}^{a}\mathcal{DO}_{ab}^{(\supR)}(r)\frac{\varpi^{2a+2b+1}}{s^{2b+2}}\nonumber\\
&+\sum_{a=1}^{m-1}\sum_{b=1}^{a}\mathcal{DO}_{ab}^{(\supP)}(r)\varpi^{2a-2b+1}s^{2b-2}\nonumber\\
 &+\sum_{a=1}^{m-2}\sum_{b=0}^{a-1}\mathcal{TO}_{ab}^{(\supR)}(r)\frac{\varpi^{2a+2b+3}}{s^{2b+2}}\nonumber\\
&+\sum_{a=0}^{m-2}\sum_{b=0}^{a}\mathcal{TO}_{ab}^{(\supL)}(r)\varpi^{2b+1}s^{2a-2b}\log(s^{2}/L^{2})\nonumber\\
&+\sum_{a=1}^{m-2}\sum_{b=0}^{a}\mathcal{TO}_{ab}^{(\supP)}(r)\varpi^{2b+1}s^{2a-2b}\Bigg\}
\nonumber \\ &  \qquad +\mathcal{O}(\epsilon^{2m}\log\epsilon).
\end{align}
\end{subequations}
In the following section, we use the expansion (\ref{eq:KHadext}) of the Hadamard parametrix to renormalize the Euclidean Green function and hence find renormalized expectation values of the scalar condensate, current and stress-energy tensor.

\section{Renormalized expectation values}
\label{sec:expectation}

Previous work \cite{Balakumar:2019djw} on the use of Hadamard renormalization for a charged scalar field outlined how to find renormalized expectation values on a Lorentzian space-time.
Since we are working on a Euclidean space-time, in this section we first outline how the results of \cite{Balakumar:2019djw} translate to a Euclidean framework.

\subsection{Euclidean vs Lorentzian expressions}
\label{sec:dictionary}

An immediate difficulty arises when one comes to computing renormalized expectation values using Euclidean methods: how do we translate the formal Lorentzian expressions for the renormalized expectation values \cite{Balakumar:2019djw} into their Euclidean counterpart? 
For a neutral scalar field on a static space-time, this is a straightforward matter since one simply replaces the Feynman Green function with the  corresponding Euclidean Green function via the mapping
\begin{align}
\label{eq:Euclideanmap}
    -iG_{\subF}(t,\mathbf{x};t',\mathbf{x}')\to G_{\subE}(\tau,\mathbf{x}; \tau', \mathbf{x}'),
\end{align}
where the two-point functions are related by
\begin{align}
   G_{\subE}(-it,\mathbf{x}; -it', \mathbf{x}') =-iG_{\subF}(t,\mathbf{x};t',\mathbf{x}').
   \label{eq:Euclideanmap1}
\end{align}
Our conventions for the Feynman Green function are given by the definition
\begin{align}
\label{eq:EuclideanFeynman}
    -iG_{\subF}(x,x')=\langle 0|\mathrm{T}\{\hat{\Phi}(x)\,\hat{\Phi}^{\dagger}(x')\}|0\rangle ,
\end{align}
where $\mathrm{T}$ denotes a time-ordering operator which permutes the field operators so that the one evaluated at a time in the chronological past of the other acts first. 

While the correspondence (\ref{eq:Euclideanmap1}) between the Feynman and Euclidean propagators remains true for a charged scalar field in static spacetimes, the Feynman propagator is a complex bidistribution and certain expectation values for the charged field involve taking real or imaginary parts of this bidistribution \cite{Balakumar:2019djw}. 
On the other hand, the Euclidean Green function (\ref{eq:GE}) is purely real, so translating the Lorentzian definitions into corresponding Euclidean ones requires consideration. As an illustrative example, consider the unrenormalized expectation value of the current, which is defined to be \cite{Balakumar:2019djw}
\begin{align}
\label{eq:current1}
    \langle \hat{J}_{\alpha }\rangle_{\textrm{unren}}=-\frac{q}{4\pi}\Big[\Im\{D_{\alpha }[-iG_{\subF}(x,x')]\}\Big],
\end{align}
where $[\cdot]$ indicates the coincidence limit $x'\to x$ and $\Im\{\cdot\}$ indicates the imaginary part. Clearly employing the mapping in (\ref{eq:Euclideanmap}) in this Lorentzian definition does not give the correct definition of the current in terms of the Euclidean Green function since the Euclidean Green function, as well as the Euclideanized version of the operator $D_{\mu}$, are real and hence its imaginary part vanishes identically. 

To obtain the correct dictionary between the Lorentzian and Euclidean definitions, one must first recast the Lorentzian definitions into an equivalent form that involves the symmetric and antisymmetric parts of the Euclidean Green function, rather than real and imaginary parts. 
In particular, because the Feynman propagator for a complex scalar field is sesquisymmetric,
\begin{align}
    G_{\subF}(x,x')=G_{\subF}^{*}(x',x),
\end{align}
this immediately implies that
\begin{align}
    \Re\{-iG_{\subF}(x,x')\}&=-iG_{\subF}^{\sy}(x,x')\nonumber\\
    i\Im\{-iG_{\subF}(x,x')\}&=-iG_{\subF}^{\asy}(x,x'),
\end{align}
Hence, we can show that an equivalent definition of the current (\ref{eq:current1}) is
\begin{align}
\label{eq:currentunren}
   \langle \hat{J}_{\alpha}\rangle_{\textrm{unren}}=\frac{q}{4\pi}\Big[\nabla_{\alpha }G_{\subF}^{\asy}(x,x')-iqA_{\alpha }G_{\subF}^{\sy}(x,x')\Big]. 
\end{align}
This definition of the current has an immediate counterpart in terms of the Euclidean Green function under the mapping
\begin{align}
\label{eq:Euclideanmap2}
    -iG_{\subF}^{\sy}(x,x')&\to G_{\subE}^{\sy}(x,x'),\nonumber\\
    -iG_{\subF}^{\asy}(x,x')&\to G_{\subE}^{\asy}(x,x').
\end{align}

This now provides the framework for defining the expectation values on the Euclidean section: we recast the Lorentzian definitions in Ref.~\cite{Balakumar:2019djw} in terms of the symmetric and antisymmetric parts of the Feynman Green function, and then map to the symmetric and antisymmetric parts of the corresponding Euclidean Green function.

\subsection{Scalar condensate}
\label{sec:condensate}

The unrenormalized scalar condensate is defined to be
\begin{align}
\label{eq:scalarcondensateunrenF}
    \langle \hat{\Phi}\hat{\Phi}^{\dagger}\rangle_{\textrm{unren}}=\Big[-iG_{\subF}(x,x')\Big],
\end{align}
and according to our prescription above, we define this in terms of the Euclidean Green function as
    \begin{align}
\label{eq:scalarcondensateunren}
    \langle \hat{\Phi}\hat{\Phi}^{\dagger}\rangle_{\textrm{unren}}=\Big[G_{\subE}(x,x')\Big].
\end{align}
Of course it is the renormalized expectation value we wish to compute and this is obtained by subtracting the Hadamard parametrix from the Euclidean Green function. 
We define the renormalized Euclidean Green function by
\begin{align}
    W(x,x')=G_{\subE}(x,x')-K(x,x').
\end{align}
It will also be useful, given our discussion in the previous subsection, to define the symmetric and antisymmetric parts
\begin{align}
\label{eq:Gren}
    W^{\sy}(x,x')&=G_{\subE}^{\sy}(x,x')-K^{\sy}(x,x'),\nonumber\\
    W^{\asy}(x,x')&=G_{\subE}^{\asy}(x,x')-K^{\asy}(x,x').
\end{align}
Now since the Hadamard parametrix, by definition, has the same short-distance singularity structure as the Euclidean Green function, $W(x,x')$ is regular in the coincidence limit $x'\to x$. 
Hence,
\begin{align} 
w(x)=\Big[W(x,x')\Big]=\Big[W^{\sy}(x,x')\Big] ,
\end{align}
where the last equality follows from the fact that the coincidence limit of a regular antisymmetric biscalar must vanish. 
Hence, we have
\begin{align}
\langle \hat{\Phi}\hat{\Phi}^{\dagger}\rangle_{\textrm{ren}}=w(x).
\end{align}

To compute $w(x)$ in practice, we require a means to express each of the terms in $K^{\sy}(x,x')$ (\ref{eq:HadamardSym}), the symmetric part of the Hadamard parametrix, as a mode-sum representation of the same form as the mode-sum representation of the Euclidean Green function (\ref{eq:GE}) for the quantum field. 
If such a mode-sum representation can be obtained, then we can renormalize the two-point function mode-by-mode. 
It is sufficient to have mode-sum representations for terms in (\ref{eq:HadamardSym}) that are nonpolynomial in $\varpi^{2}$ and $s^{2}$, since the terms that are polynomial in these variables have only finitely many modes in their mode sum representations and hence cannot affect the large $n$, large $l$ behaviour. 
Hence, only terms of the form $\varpi^{2a+2b}/s^{2b+2}$ and $\varpi^{2b}s^{2a-2b}\log(s^{2})$ require a mode-sum representation. 
Formally speaking, some of these terms are vanishing in the coincidence limit but their inclusion as a mode-sum serves to increase the rate of convergence of the mode-sum representation of the renormalized expectation value. 
For example, terms of the form $\varpi^{2a+2b}/s^{2b+2}\sim\mathcal{O}(\Delta x^{2a-2})$ with $a\ge 2$ are formally vanishing in the coincidence limit, but by including them as a mode-sum we have complete control of the rate of convergence of the mode-sum. 

To obtain the mode-sum representation of terms of the form $\varpi^{2a+2b}/s^{2b+2}$, we start with the mode-sum ansatz \cite{Taylor:2016edd,Taylor:2017sux,Taylor:2022sly}
\begin{align}
\label{eq:modesumansatz}
    \frac{\varpi^{2a+2b}}{s^{2b+2}}=\sum_{l=0}^{\infty}(2l+1)P_{l}(\cos\gamma)\sum_{n=-\infty}^{\infty}e^{in\kappa\Delta\tau}\Psi_{nl}(a,b|r),
\end{align}
for some as-yet undetermined functions $\Psi_{nl}(a,b|r)$ which we call the regularization parameters. 
This ansatz (\ref{eq:modesumansatz}) can be inverted using the completeness relations of the Legendre polynomials and the Fourier modes to obtain a double integral expression for the regularization parameters. 
Note that the fact that the point-splitting in $K^{\sy}(x,x')$ was maintained in multiple directions (we have taken only radial points together $r'\to r$) played an essential role in this step. 
Moreover, the fact that we expanded using the ``extended coordinates'' $\varpi$ and $s$ (\ref{eq:extendedcoords}) makes it feasible to compute the double integral in terms of known functions. 
The result depends only on the geometry of space-time and hence is the same in the charged case as in the neutral case. We state only the result, further details can be found in Refs.~\cite{Taylor:2016edd,Taylor:2017sux,Taylor:2022sly}:
\begin{multline}
\label{eq:Psi}
	\Psi_{n l}(a,b|r)=\frac{2^{a-b}a!(2a-1)!!(-1)^{n}}{\kappa^{2a+2b}r^{2b+2}b!}
	\\ \times\sum_{p=n-a}^{n+a}\left(\frac{1}{\eta}\frac{\partial}{\partial\eta}\right)^{b}\frac{(-1)^{b}P_{l}^{-|p|}(\eta)Q_{l}^{|p|}(\eta)}{(a-n+p)!(a+n-p)!},
\end{multline}
with
\begin{align}
\eta=\sqrt{1+\frac{f(r)}{\kappa^{2}r^{2}}},
\end{align}
and where $P_{\nu}^{\mu}(z)$, $Q_{\nu}^{\mu}(z)$ are the associated Legendre functions of the first and second kind, respectively. It is straightforward to show that $\Psi_{nl}(a,b|r)$ is invariant under the mapping $n\to -n$ so that one could equally write
\begin{multline}
\frac{\varpi^{2a+2b}}{s^{2b+2}} \\ =\sum_{l=0}^{\infty}(2l+1)P_{l}(\cos\gamma)\sum_{n=-\infty}^{\infty}\cos(n\kappa\Delta\tau)\Psi_{nl}(a,b|r),
\end{multline}
which is precisely of the same form as (\ref{eq:GESym}). The rational terms coming from the tail part of the Hadamard parametrix are of the form $\varpi^{2a+2b+2}/s^{2b+2}$ which are trivially obtained from those above with $a\to a+1$.

All that remains is to express the logarithmic terms $\varpi^{2b}s^{2a-2b}\log(s^{2})$ in (\ref{eq:HadamardSym}) as a mode-sum. 
The approach is very similar and indeed the regularization parameters themselves are essentially identical to the neutral scalar case \cite{Taylor:2017sux,Taylor:2022sly}. 
We have
\begin{multline}
    \varpi^{2b}s^{2a-2b}\log(s^{2}) \\ =\sum_{l=0}^{\infty}(2l+1)P_{l}(\cos\gamma)
    \sum_{n=-\infty}^{\infty}\cos(n\kappa\Delta\tau)\chi_{nl}(a,b|r),
\end{multline}
where
\begin{widetext}
\begin{align}
    \chi_{nl}(a,b|r)=\begin{cases}
 		\displaystyle{	\frac{(-1)^{n}(a-b)!(2b)!}{2\kappa^{2b}r^{2b-2a}}\sum_{k=0}^{1+a-b}(-1)^{k}\binom{1+a-b}{k}\frac{(l+\tfrac{3}{2}+a-b-2k)}{(l+\tfrac{1}{2}-k)_{2+a-b}}	}
   \\ \hspace{6cm} \times 
  \displaystyle{ \sum_{p=n-b}^{n+b}\frac{P_{l+a-b+1-2k}^{-|p|}(\eta)Q_{l+a-b+1-2k}^{|p|}(\eta)}{(b-n+p)!(b+n-p)!}} \\ \qquad\qquad\qquad\qquad\qquad\qquad\qquad\qquad\qquad\qquad\qquad\qquad\qquad\qquad\qquad\qquad\qquad\textrm{for}\,\,\,\, l>a-b, \\ \null \\
 		\displaystyle{\frac{2^{b-1}(-1)^{l}}{\pi \kappa^{2b-1}}\left[\frac{d}{d\lambda}(\lambda+1-l)_{l}(2r^{2})^{\lambda}\int_{0}^{2\pi/\kappa}(1-\cos\kappa t)^{b}e^{-in\kappa t}(z^{2}-1)^{(\lambda+1)/2}\mathcal{Q}_{l}^{-\lambda-1}(z) \, dt\right]_{\lambda=a-b}} \\		\qquad\qquad\qquad\qquad\qquad\qquad\qquad\qquad\qquad\qquad\qquad\qquad\qquad\qquad\qquad\qquad\qquad\textrm{for}\,\,\,\,	l\le a-b,
 	\end{cases}
  \label{eq:chi}
\end{align}
\end{widetext}
where 
\begin{equation}
z=\eta^{2}-(\eta^{2}-1)\cos(\kappa t).
\end{equation} 
Again, $P^{\mu}_{\nu}(z)$, $Q^{\mu}_{\nu}(z)$ are the associated Legendre functions of the first and second kind, respectively, while $\mathcal{Q}^{\mu}_{\nu}(z)$ is Olver's definition \cite{NIST:DLMF} of the Legendre function of the second kind. 
The integral in the second branch of the expression above (\ref{eq:chi}) can be obtained in closed form in terms of special functions, but the expression is lengthy and offers no significant computational advantage compared to simply computing the integral above numerically. 
Besides, this expression is only required for the very low $l$ modes with $l\le a-b$.

Putting all of these details together then, we obtain the following expression for the renormalized scalar condensate for a charged scalar field in a thermal state
\begin{align} 
\label{eq:rencondensate}
\langle\hat{\Phi}\hat{\Phi}^{\dagger}\rangle_{\textrm{ren}}=\frac{1}{8\pi^{2}}\sum_{l=0}^{\infty}(2l+1)\sum_{n=-\infty}^{\infty}\{g_{nl}(r)-k_{nl}^{(m)}(r)\}\nonumber\\
    -\frac{1}{8\pi^{2}}\mathcal{DE}^{(\supP)}_{11}(r)+\frac{1}{8\pi^{2}}\mathcal{TE}^{(\supL)}_{00}\log(L^{2}),
\end{align}
where
\begin{align}
\label{eq:regmodes}
    k_{nl}^{(m)}(r)   = & \sum_{a=0}^{m}\sum_{b=0}^{a}\mathcal{DE}^{(\supR)}_{ab}(r)\,\Psi_{nl}(a,b|r)\nonumber\\
    & +\sum_{a=1}^{m-1}\sum_{b=0}^{a-1}\mathcal{TE}^{(\supR)}_{ab}(r)\Psi_{nl}(a+1,b|r)\nonumber\\
    & +\sum_{a=0}^{m-1}\sum_{b=0}^{a}\mathcal{TE}_{ab}^{(\supL)}(r)\chi_{nl}(a,b|r).
\end{align}
The renormalized scalar condensate (\ref{eq:rencondensate}) is real, since $g_{nl}(r) = g_{nl}(r,r)$ satisfies (\ref{eq:gnpm}) and $k_{nl}^{(m)}(r)$ is real.
In principle, we can make the mode sum in (\ref{eq:rencondensate}) increasingly convergent by choosing $m$ larger and larger. Having this precise control over the rate of convergence with a direct correspondence to the order of expansion in the Hadamard parametrix is one of the major advantages of this approach.

\subsection{Renormalized current}
\label{sec:current}

We now turn to the calculation of the renormalized current $\langle \hat{J}_{\alpha }\rangle_{\textrm{ren}}$. We have already derived, in Sec.~\ref{sec:dictionary}, the unrenormalized current in terms of the symmetric and antisymmetric parts of the Feynman propagator, given in Eq.~(\ref{eq:currentunren}). 
In terms of the Euclidean Green function, we employ the mapping (\ref{eq:Euclideanmap2}) and the renormalization (\ref{eq:Gren}) to obtain
\begin{align}
    \langle \hat{J}_{\alpha }\rangle_{\textrm{ren}}=\frac{iq}{4\pi}\Big[\nabla_{\alpha }W^{\asy}(x,x')-iqA_{\alpha }\,W^{\sy}(x,x')\Big].
\end{align}
Moreover, we can expand the biscalar $W(x,x')$ in a covariant Taylor expansion about the point $x$: 
\begin{align}  
W(x,x')=w(x)+w_{\alpha}(x)\sigma^{\alpha}+\tfrac{1}{2}w_{\alpha\beta}(x)\sigma^{\alpha}\sigma^{\beta}+\mathcal{O}(\epsilon^{3}),
\end{align}
where $w_{\alpha}$ and $w_{\alpha\beta}$ are tensors at the point $x$,  
and, as before, we use the book-keeping notation $\mathcal{O}(\epsilon)\sim\mathcal{O}(\Delta x)$ to track orders in a short-distance expansion. 
A similar expansion for $W(x',x)$ is obtained by swapping $x$ and $x'$ above. 
It is then straightforward to show that
\begin{align}
    w_{\alpha}&=\left[\nabla_{\alpha}W^{\asy}(x,x')\right]-\tfrac{1}{2}\nabla_{\alpha}w,\nonumber\\
    w_{(\alpha\beta)}&=\tfrac{1}{2}w_{;\alpha\beta}+w_{(\alpha;\beta)}-\left[\nabla_{\beta}\nabla_{\alpha}W^{\sy}(x,x')\right],\nonumber\\
    &=w_{(\alpha;\beta)}+\left[\nabla_{\beta'}\nabla_{\alpha}W^{\sy}(x,x')\right],
\end{align}
where the last equality follows from Synge's Rule and the fact that $[\nabla_{\alpha}W^{\sy}(x,x')]=\tfrac{1}{2}\nabla_{\alpha}w$. 
We can then rewrite the renormalized current as
\begin{align}  
\langle\hat{J}_{\alpha }\rangle_{\textrm{ren}}=\frac{iq}{4\pi}\Big(w_{\alpha }+\tfrac{1}{2}\nabla_{\alpha }w-iqA_{\alpha }w\Big).
\end{align}

Now the only nonzero component of the current for a charged scalar field in the Hartle-Hawking state on a static spacetime with an electrostatic potential is the time component, which is
\begin{align}
\label{eq:Jtau}
\langle \hat{J}_{\tau}\rangle_{\textrm{ren}}=\frac{iq}{4\pi}\Big(w_{\tau}-iqA_{\tau}w\Big),
\end{align}
where we have discussed already how to compute $w=\langle \hat{\Phi}\hat{\Phi}^{\dagger}\rangle_{\textrm{ren}}$ and 
\begin{align}   w_{\tau}=\Big[\partial_{\tau}G_{\subE}^{\asy}(x,x')-\partial_{\tau}K^{\asy}(x,x')\Big].
\end{align}
The first term is straightforward to compute mode-by-mode and is given by
\begin{multline}    
\label{eq:Gasymtau}
\partial_{\tau}G_{\subE}^{\asy}(x,x')
\\ =\frac{i}{8\pi^{2}}\sum_{l=0}^{\infty}P_{l}(\cos\gamma)\sum_{n=-\infty}^{\infty}(n\kappa)\cos(n\kappa\Delta\tau)g_{nl}(r,r').
\end{multline}

To renormalize this mode-by-mode, we require a mode-sum representation for $\partial_{\tau}K^{\asy}(x,x')$. 
The regularization parameters that we derived for the terms in the symmetric part of the parametrix (\ref{eq:HadamardSym}) are not valid for the antisymmetric part (\ref{eq:HadamardASym}), since the time integral involved in inverting the mode-sum ansatz is harder to express in closed form when there are odd powers of $\varpi$ in the integral. 
Nevertheless, we can proceed as follows: Differentiating $K^{\asy}(x,x')$ (\ref{eq:HadamardASym}) with respect to $\tau$ using the definitions of $\varpi$ and $s$ (\ref{eq:extendedcoords}) gives
\begin{widetext}
 \begin{align}
     \partial_{\tau}K^{\asy}(x,x')= & \frac{1}{8\pi^{2}}\Bigg\{\sum_{a=0}^{m-1}\sum_{b=0}^{a}\mathcal{DO}_{ab}^{(\supR)}(r)\frac{\varpi^{2a+2b}}{s^{2b+2}}\left[(2a+2b+1)-(2b+2)f(r)\frac{\varpi^{2}}{s^{2}}\right]\frac{d\varpi}{d\tau}\nonumber\\
&+\sum_{a=1}^{m-1}\sum_{b=1}^{a}\mathcal{DO}_{ab}^{(\supP)}(r)\varpi^{2a-2b}s^{2b-2}\left[(2a-2b+1)+(2b-2)f(r)\frac{\varpi^{2}}{s^{2}}\right]\frac{d\varpi}{d\tau}\nonumber\\
 &+\sum_{a=1}^{m-2}\sum_{b=0}^{a-1}\mathcal{TO}_{ab}^{(\supR)}(r)\frac{\varpi^{2a+2b+2}}{s^{2b+2}}\left[(2a+2b+3)-(2b+2)f(r)\frac{\varpi^{2}}{s^{2}}\right]\frac{d\varpi}{d\tau}\nonumber\\
&+\sum_{a=0}^{m-2}\sum_{b=0}^{a}\mathcal{TO}_{ab}^{(\supL)}(r)\varpi^{2b}s^{2a-2b}\left[(2b+1)\log(s^{2}/L^{2})+(2a-2b)\frac{\varpi^{2}}{s^{2}}f(r)\,\log(s^{2}/L^{2})+2 f(r)\frac{\varpi^{2}}{s^{2}}\right]\frac{d\varpi}{d\tau}\nonumber\\
&+\sum_{a=1}^{m-2}\sum_{b=0}^{a}\mathcal{TO}_{ab}^{(\supP)}(r)\varpi^{2b}s^{2a-2b}\left[(2b+1)+(2a-2b)f(r)\frac{\varpi^{2}}{s^{2}}\right]\frac{d\varpi}{d\tau}\Bigg\}
\nonumber \\ & \qquad +\mathcal{O}(\epsilon^{2m-1}\log\epsilon) .
 \end{align}   
We then note that
\begin{equation}
    \frac{d\varpi}{d\tau}=\sum_{k=0}^{\infty}\binom{\frac{1}{2}}{k}\left(-\frac{\kappa^{2}\varpi^{2}}{4}\right)^{k},
\end{equation}
which involves only even powers of $\varpi$. 
Moreover, we need not worry that this is an infinite sum since we can truncate at the appropriate order consistent with the order in which we are expanding the Hadamard parametrix. 
We thus obtain
\begin{align}
     \partial_{\tau}K^{\asy}(x,x')= &\frac{1}{8\pi^{2}}\Bigg\{\sum_{a=0}^{m-1}\sum_{b=0}^{a}\sum_{k=0}^{m-a}\binom{\frac{1}{2}}{k}\left(-\frac{\kappa^{2}}{4}\right)^{k}\mathcal{DO}_{ab}^{(\supR)}(r)\frac{\varpi^{2a+2b+2k}}{s^{2b+2}}\left[(2a+2b+1)-(2b+2)f(r)\frac{\varpi^{2}}{s^{2}}\right]\nonumber\\
&+\sum_{a=1}^{m-1}\sum_{b=1}^{a}\sum_{k=0}^{m-a}\binom{\frac{1}{2}}{k}\left(-\frac{\kappa^{2}}{4}\right)^{k}\mathcal{DO}_{ab}^{(\supP)}(r)\varpi^{2a-2b+2k}s^{2b-2}\left[(2a-2b+1)+(2b-2)f(r)\frac{\varpi^{2}}{s^{2}}\right]\nonumber\\
 &+\sum_{a=1}^{m-2}\sum_{b=0}^{a-1}\sum_{k=0}^{m-a-2}\binom{\frac{1}{2}}{k}\left(-\frac{\kappa^{2}}{4}\right)^{k}\mathcal{TO}_{ab}^{(\supR)}(r)\frac{\varpi^{2a+2b+2k+2}}{s^{2b+2}}\left[(2a+2b+3)-(2b+2)f(r)\frac{\varpi^{2}}{s^{2}}\right]\nonumber\\
&+\sum_{a=0}^{m-2}\sum_{b=0}^{a}\sum_{k=0}^{m-a-2}\binom{\frac{1}{2}}{k}\left(-\frac{\kappa^{2}}{4}\right)^{k}\mathcal{TO}_{ab}^{(\supL)}(r)\varpi^{2b+2k}s^{2a-2b}\bigg[(2b+1)\log(s^{2}/L^{2})\nonumber\\
&\qquad\qquad\qquad\qquad\qquad\qquad\qquad\qquad\qquad+(2a-2b)\frac{\varpi^{2}}{s^{2}}f(r)\,\log(s^{2}/L^{2})+2 f(r)\frac{\varpi^{2}}{s^{2}}\bigg]\nonumber\\
&+\sum_{a=1}^{m-2}\sum_{b=0}^{a}\sum_{k=0}^{m-a-2}\binom{\frac{1}{2}}{k}\left(-\frac{\kappa^{2}}{4}\right)^{k}\mathcal{TO}_{ab}^{(\supP)}(r)\varpi^{2b+2k}s^{2a-2b}\left[(2b+1)+(2a-2b)f(r)\frac{\varpi^{2}}{s^{2}}\right]\Bigg\}
\nonumber \\ & \qquad  +\mathcal{O}(\epsilon^{2m-1}\log\epsilon).
\label{eq:Kafinal}
 \end{align}  
Now all of the terms in the expression (\ref{eq:Kafinal}) can be expressed as a mode-sum using the regularization parameters $\Psi_{nl}(a,b|r)$ (\ref{eq:Psi}) and $\chi_{nl}(a,b|r)$ (\ref{eq:chi}). 
As before, terms that are polynomial in both $\varpi^{2}$ and $s^{2}$ do not require a mode-sum representation, whilst terms that are nonpolynomial are converted to a mode-sum even if they formally vanish in the coincidence limit. 
Hence we have 
\begin{subequations}
\begin{align}
\label{eq:wtau}
    w_{\tau}=&\frac{1}{8\pi^{2}}\sum_{l=0}^{\infty}(2l+1)\sum_{n=-\infty}^{\infty} \{ i(n\kappa)g_{nl}(r)-j_{nl}^{(m)}(r)\}
    -\frac{1}{8\pi^{2}}\mathcal{DO}_{11}^{(\supP)}(r) +\frac{1}{8\pi^{2}}\mathcal{TO}_{00}^{(\supL)}(r)\log(L^{2}),
\end{align}
where
\begin{align}
\label{eq:jm}
    j_{nl}^{(m)}(r)  = &\sum_{a=0}^{m-1}\sum_{b=0}^{a}\sum_{k=0}^{m-a}\binom{\tfrac{1}{2}}{k}\left(-\frac{\kappa^{2}}{4}\right)^{k}\mathcal{DO}_{ab}^{(\supR)}(r)\Big[(2a+2b+1)\Psi_{nl}(a+k,b|r)-(2b+2)f(r)\Psi_{nl}(a+k,b+1|r)\Big]\nonumber\\
    & + \sum_{a=1}^{m-2}\sum_{b=0}^{a-1}\sum_{k=0}^{m-a-2}\binom{\tfrac{1}{2}}{k}\left(-\frac{\kappa^{2}}{4}\right)^{k}\mathcal{TO}_{ab}^{(\supR)}(r)\Big[(2a+2b+3)\Psi_{nl}(a+k+1,b|r)
    \nonumber \\ & \qquad\qquad\qquad\qquad\qquad\qquad\qquad\qquad\qquad\qquad\qquad\qquad\qquad
    -(2b+2)f(r)\Psi_{nl}(a+k+1,b+1|r)\Big]\nonumber\\
    & + \sum_{a=0}^{m-2}\sum_{b=0}^{a}\sum_{k=0}^{m-a-2}\binom{\tfrac{1}{2}}{k}\left(-\frac{\kappa^{2}}{4}\right)^{k}\mathcal{TO}_{ab}^{(\supL)}(r)\Big[(2b+1)\chi_{nl}(a,b+k|r)+(2a-2b)f(r)\chi_{nl}(a-1,b+k+1|r)\Big]\nonumber\\
    & + \sum_{a=0}^{m-2}\sum_{k=0}^{m-a-2}\binom{\tfrac{1}{2}}{k}\left(-\frac{\kappa^{2}}{4}\right)^{k}\mathcal{TO}_{aa}^{(\supL)}(r)\Big[2f(r)\Psi_{nl}(a+k+1,0|r)\Big].
\end{align}
\end{subequations}
\end{widetext}
We now have convergent mode-sum expressions for $w$ and $w_{\tau}$ which are combined according to (\ref{eq:Jtau}) to give $\langle \hat{J}_{\tau}\rangle_{\textrm{ren}}$. 
Finally, since the Euclidean time and Lorentzian time are related by the mapping $t\to-i\tau$, we have
\begin{align} 
\label{eq:Jt}
\langle\hat{J}_{t}\rangle_{\textrm{ren}}=i\langle\hat{J}_{\tau}\rangle_{\textrm{ren}}.
\end{align}
Note that, from (\ref{eq:Jtau}), since $A_{\tau }$ is purely imaginary, so too is $\langle\hat{J}_{\tau}\rangle_{\textrm{ren}}$.
Hence $\langle\hat{J}_{t}\rangle_{\textrm{ren}}$ (\ref{eq:Jt}) is real, as expected.

\subsection{Renormalized stress energy tensor}
\label{sec:SET}

The formal expression for the unrenormalized stress-energy tensor on the Lorentzian space-time is \cite{Balakumar:2019djw}
\begin{align}
    \langle \hat{T}^{\alpha  }{}_{\beta}\rangle_{\textrm{unren}}
    =\Big[\Re\left\{\hat{\mathsf{T}}^{\alpha }{}_{\beta }(-iG_{\subF}(x,x'))\right\}\Big]
    +\frac{1}{4\pi^{2}}\delta^{\alpha }{}_{\beta }v_{1},
\end{align}
where $\hat{\mathsf{T}}^{\alpha}{}_{\beta }$ is the differential operator
\begin{align}
    \hat{\mathsf{T}}^{\alpha}{}_{\beta}  = & 
    (1-2\xi)g_{\beta}{}^{\lambda '}D^{\alpha  }D_{\lambda'}^{*}
    +(2\xi-\tfrac{1}{2})\delta^{\alpha }{}_{\beta }g^{\rho\lambda'}D_{\rho}D_{\lambda'}^{*}
    \nonumber\\
& -2\xi\,D^{\alpha }D_{\beta}
+2\xi\,\delta^{\alpha }{}_{\beta }D_{\rho}D^{\rho}+\xi \,R^{\alpha }{}_{\beta }
\nonumber \\ & -\tfrac{1}{2}(\mu ^{2}+\xi\,R)\delta^{\alpha  }{}_{\beta },
\end{align}
with $g^{\rho \lambda '}$ the bivector of parallel transport, and
\begin{align}
    v_{1}=&
    \tfrac{1}{8}\left[ \mu^{2}+(\xi-\tfrac{1}{6})R\right]^{2}
    -\tfrac{1}{24}(\xi-\tfrac{1}{5})\Box R
    -\tfrac{1}{720}R^{\alpha\beta}R_{\alpha \beta }\nonumber\\
    &+\tfrac{1}{720}R^{\alpha \beta \rho\lambda}R_{\alpha \beta \rho\lambda}
    -\tfrac{1}{48}q^{2}F^{\alpha \beta }F_{\alpha \beta }
\end{align}
is the local term which must be added to ensure the stress-energy tensor is conserved. 
The derivatives and the electromagnetic potential here are to be understood as the Lorentzian quantities. 

We can follow the procedure from Sec.~\ref{sec:dictionary} and express the unrenormalized stress-energy tensor in terms of derivatives acting on the symmetric and antisymmetric parts of the Feynman Green function, which is then easily mapped to the Euclidean section via (\ref{eq:Euclideanmap2}). 
Renormalizing the Euclidean Green function then gives the following definition of the RSET
\begin{align}
\label{eq:rset1}
   \langle \hat{T}^{\alpha  }{}_{\beta }\rangle_{\textrm{ren}} = &  
   -\Big[\left(\nabla^{\alpha }\nabla_{\beta }
   -\tfrac{1}{2}\delta^{\alpha }{}_{\beta }\Box \right)W^{\sy}(x,x')\Big]
   \nonumber\\
 &  + \Big\{-(\xi-\tfrac{1}{2})\nabla^{\alpha }\nabla_{\beta }
 +(\xi-\tfrac{1}{4})\delta^{\alpha }{}_{\beta }\,\Box
  \nonumber\\
 & +q^{2}A^{\alpha  }A_{\beta }
   -\tfrac{1}{2}q^{2}A^{\lambda}A_{\lambda}\delta^{\alpha }{}_{\beta }
   -\tfrac{1}{2}(\mu ^{2}+\xi\,R)\delta^{\alpha }{}_{\beta }
   \nonumber\\
 &  +\xi R^{\alpha  }{}_{\beta } +iq\left(A^{(\alpha }\nabla_{\beta )}
 -\tfrac{1}{2}\delta^{\alpha }{}_{\beta }A^{\lambda}\nabla_{\lambda}\right)\Big\}w
 \nonumber\\
  &  +i q\left(2 A^{(\alpha }w_{\beta )}-\delta^{\alpha }{}_{\beta }A^{\lambda}w_{\lambda}\right) 
  +\frac{1}{4\pi^{2}}\delta^{\alpha }{}_{\beta }v_{1},
\end{align}
where round brackets surrounding indices denotes symmetrization.
In arriving at this expression, we have used Synge's rule for exchanging limits and derivatives, as well as the fact that the coincidence limit of the antisymmetric part of the propagator vanishes. 
In this expression, all quantities are to be thought of as Euclidean so, for example, the d'Alembertian operator $\Box=\nabla_{\mu}\nabla^{\mu}$ is with respect to the Euclidean metric. 

We have already discussed how to compute $w$ and $w_{\alpha }$ in Secs.~\ref{sec:condensate} and \ref{sec:current} respectively. 
Moreover, the derivatives of these quantities that appear in Eq.~(\ref{eq:rset1}) are straightforward, since only radial derivatives are nonvanishing and these can be accurately obtained numerically, provided that $w$ and $w_{\alpha }$ are computed with sufficient accuracy on a sufficiently fine radial grid. 

This leaves only the first term in (\ref{eq:rset1}), which we now consider, 
following the approach of \cite{Taylor:2022sly}. 
We note that the Euclidean Green function $G_{\subE}(x,x')$ satisfies the charged scalar field equation (\ref{eq:waveeqn}) with the vanishing right-hand-side replaced by a delta function, so that the renormalized Green function $W(x,x')$ satisfies the inhomogeneous wave equation with a regular source. 
In the coincidence limit, this gives
\begin{align}
    \Big[\left(D_{\alpha }D^{\alpha }-\mu ^{2}-\xi R\right)W(x,x')\Big]=-\frac{3}{4\pi^{2}}v_{1}.
\end{align}
Moreover, considering the symmetric and antisymmetric parts gives the coincidence limit
\begin{align}
\label{eq:waveW}
\Big[\Box W^{\sy}(x,x')\Big]&=(m^{2}+\xi R+q^{2}A_{\lambda}A^{\lambda}+i q A^{\lambda}\nabla_{\lambda})w\nonumber\\
& \qquad +2i q A^{\lambda}w_{\lambda}-\frac{3}{4\pi^{2}}v_{1},
\end{align}
and employing this result in (\ref{eq:rset1}) gives
\begin{multline}
\label{eq:rset2}
   \langle \hat{T}^{\alpha }{}_{\beta }\rangle_{\textrm{ren}}
   =  -\widetilde{w}^{\alpha }{}_{\beta } 
   \\
   + \Big\{-(\xi-\tfrac{1}{2})\nabla^{\alpha }\nabla_{\beta }
   +(\xi-\tfrac{1}{4})\delta^{\alpha }{}_{\beta }\,\Box
   \\
   +q^{2}A^{\alpha }A_{\beta } 
   +\xi R^{\alpha }{}_{\beta } +iq A^{(\alpha }\nabla_{\beta )}\Big\}w
   \\
   +2iq A^{(\alpha }w_{\beta )} -\frac{1}{8\pi^{2}}\delta^{\alpha }{}_{\beta }v_{1},
\end{multline}
where we have introduced the notation 
\begin{align}
    \widetilde{w}^{\alpha }{}_{\beta }=\Big[\nabla^{\alpha }\nabla_{\beta }W^{\sy}(x,x')\Big].
\end{align}
For the component $\widetilde{w}^{\tau}{}_{\tau}$, we can first write
\begin{align}
    \widetilde{w}^{\tau}{}_{\tau}=-\Big[\nabla^{\tau'}\nabla_{\tau}W^{\sy}(x,x')\Big]+\frac{f_{,r}}{4}w_{,r},
\end{align}
using Synge's rule and the fact that $w$ is independent of time. The last term on the right-hand side we can compute by simply taking a radial derivative of the interpolation function for $w$. The salient point is that now the covariant derivatives in the first term on the right-hand side are at different space-time points and so act as partial derivatives.
This can now be computed from taking time derivatives of the mode-sum representation of $W^{\sy}(x,x')$ already derived (\ref{eq:rencondensate}). 
Explicitly, we have
\begin{align}
\label{eq:wtautau}
    \Big[\nabla^{\tau'}\nabla_{\tau}W^{\sy}(x,x')\Big] \hspace{-3cm} & \nonumber\\
    = & ~ \frac{1}{8\pi^{2}}\sum_{l=0}^{\infty}(2l+1)\sum_{n=-\infty}^{\infty}\frac{n^{2}\kappa^{2}}{f(r)}\left\{ g_{nl}(r)-k_{nl}^{(m)}(r)\right\} \nonumber\\
    & +\frac{1}{4\pi^{2}f(r)}\Big\{\mathcal{DE}_{21}^{(\supP)}(r)+f(r)\mathcal{DE}_{22}^{(\supP)}(r)\nonumber\\
    &-f(r)\mathcal{TE}_{10}^{(\supL)}(r)\log(L^{2})
    -\mathcal{TE}_{11}^{(\supP)}(r)\log(L^{2})\nonumber\\
    & +f(r)\mathcal{TE}_{10}^{(\supP)}(r)+\mathcal{TE}_{11}^{(\supP)}(r)\Big\}.
\end{align}
The mode-sum in the first line remains rapidly convergent for $m$ sufficiently large. The regularization modes $k_{nl}^{(m)}(r)$ are given already in (\ref{eq:regmodes}).

The angular components are found in a similar way, noting first that we have
\begin{equation}
    \widetilde{w}^{\phi}{}_{\phi}=\widetilde{w}^{\theta}{}_{\theta}=-\Big[\nabla^{\phi'}\nabla_{\phi}W^{\sy}(x,x')\Big]+\frac{f(r)}{2r}w_{,r},
\end{equation}
where the first term on the right-hand side is obtained by taking angular partial derivatives of the mode-sum representation for $W^{\sy}(x,x')$, yielding
\begin{align}
\label{eq:wphiphi1}   &\Big[\nabla^{\phi'}\nabla_{\phi}W^{\sy}(x,x')\Big]
    \nonumber\\
    \,\,= & ~\frac{1}{8\pi^{2}}\sum_{l=0}^{\infty}\frac{l(l+1)(2l+1)}{2r^{2}}\sum_{n=-\infty}^{\infty}\left\{ g_{nl}(r)-k_{nl}^{(m)}(r)\right\} 
    \nonumber\\
   &\,\, +\frac{1}{4\pi^{2}}\Big\{\mathcal{DE}_{22}^{(\supP)}(r)-\mathcal{TE}_{10}^{(\supL)}(r)\log(L^{2})+\mathcal{TE}_{10}^{(\supP)}(r)\Big\}.
\end{align}
Finally, we require a means to find $\widetilde{w}^{r}{}_{r}$. 
The most efficient way to compute this is to note that
\begin{align}
    \widetilde{w}^{r}{}_{r}=\left[\Box W^{\sy}(x,x')\right]-\widetilde{w}^{\tau}{}_{\tau}-2\widetilde{w}^{\phi}{}_{\phi},
\end{align}
and hence from (\ref{eq:waveW}), we simply have
\begin{align}
    \widetilde{w}^{r}{}_{r}=& ~-\widetilde{w}^{\tau}{}_{\tau}-2\widetilde{w}^{\phi}{}_{\phi}+2i q A^{\lambda}w_{\lambda}
    \nonumber\\
   & +(m^{2}+\xi R+q^{2}A_{\lambda}A^{\lambda}+i q A^{\lambda}\nabla_{\lambda})w
-\frac{3}{4\pi^{2}}v_{1},
\end{align}
where a mode-sum representation for every term on the right-hand side has already been developed.

\section{Hartle-Hawking state on Reissner-Nordstr\"om}
\label{sec:RNHH}
 
In this section we demonstrate the utility of the above framework by applying it to a charged quantum scalar field on the Reissner-Nordstr\"om spacetime. 
Our geometry is once again described by the line element  (\ref{eq:metric})
with
\begin{equation}
    f(r)=\left(1-\frac{2M}{r}+\frac{Q^{2}}{r^{2}}\right),
\end{equation}
where $M$ is the black hole mass and $Q$ its electric charge. 
The roots of $f(r)$ are given by 
\begin{equation}
r_{\pm } = M \pm {\sqrt {M^{2}-Q^{2}}},
\label{eq:rplus}
\end{equation}
and correspond to the event ($r_{+}$) and Cauchy ($r_{-}$) horizons. 
We take $\tau$ to be periodic with $\tau=\tau+2\pi/\kappa$, where $\kappa$ is the event horizon surface gravity, given by
\begin{equation}
    \kappa=\frac{r_{+}-r_{-}}{2r_{+}^{2}}.
\end{equation}
Therefore the charged scalar field is assumed to be in the Hartle-Hawking state \cite{Hartle:1976tp}, 
so we are considering a thermal state at the black hole temperature $\kappa /2\pi $.
The gauge field takes the form
\begin{equation}
\label{eq:gauge}
    A=\left(-\frac{Q}{r}+\frac{Q}{r_{+}}\right)dt,
\end{equation}
where we have chosen a gauge such that $A$ is regular throughout the Euclidean section.
In particular, regularity requires that we use a gauge in which $A$ vanishes on the horizon \cite{Braden:1990hw}.

\subsection{Radial modes}
\label{sec:modes}

The Euclidean Green function corresponding to the field in the Hartle-Hawking state can again be expanded as (\ref{eq:GE}), and further separated into its symmetric and antisymmetric parts, as defined in (\ref{eq:GESym}) and (\ref{eq:GEASym}) respectively.  
The mode functions $p_{nl}(r)$ and $q_{nl}(r)$ now satisfy the radial equation
\begin{align}
\label{eq:RNRad}
     \Bigg[\frac{d}{dr}\left(r^{2}f(r)\frac{d}{dr}\right)&-\frac{r^{2}}{f(r)}\left(n\kappa-iq\left\{ \frac{Q}{r}-\frac{Q}{r_{+}}\right\} \right)^{2}
    \nonumber \\ 
    -r^{2}\mu^{2}&-l(l+1)\Bigg]Y_{nl}(r)=0,
\end{align}
with $p_{nl}(r)$ regular at $r_{+}$, and $q_{nl}(r)$  regular as $r\to\infty$. 

Following a procedure analogous to that of \cite{Vieira:2021nha}, one can transform the radial equation into the confluent Heun equation \cite{Ronveaux:1995}. 
This allows one to obtain an expression for $p_{nl}(r)$ in terms of a confluent Heun function:
\begin{align}
    p_{nl}(r)= & ~e^{\frac{\alpha_{1}\left(r-r_{-}\right)}{r_{+}-r_{-}}}
    \left(\frac{r-r_{-}}{r_{+}-r_{-}}\right)^{\alpha_{2}}
    \left(\frac{r-r_{+}}{r_{+}-r_{-}}\right)^{\alpha_{3}} 
    \nonumber\\
    &\times\textsc{HeunC}\left(a_{1},a_{2},a_{3},a_{4},a_{5};\frac{r-r_{+}}{r_{-}-r_{+}}\right),
    \label{eq:pHeun}
\end{align}
where $\textsc{HeunC}\left(...\right)$ denotes the confluent Heun function satisfying the equation
\begin{multline}
    \Big[ z(z-1)\frac{d^{2}}{dz^{2}}+\left\{ a_{3}(z-1)+a_{4}z+z(z-1)a_{5}\right\} \frac{d}{dz} 
    \\
    +\left(a_{2}z-a_{1}\right) \Big]\textsc{HeunC}\left(a_{1},a_{2},a_{3},a_{4},a_{5};z\right) =0,
\end{multline}
with $z=(r-r_{+})/(r_{-}-r_{+})$, 
such that 
\begin{equation}
\textsc{HeunC}\left( a_{1},a_{2},a_{3},a_{4},a_{5};0\right)=1.
\end{equation} 
The constants in (\ref{eq:pHeun}) are given by
\begin{align}
    \alpha_{1} =& ~\frac{r_{+}-r_{-}}{r_{+}} {\sqrt{r_{+}^2 \left(\mu ^2+\kappa ^2 n^2\right)+2 i \kappa  n q Q r_{+}-q^2 Q^2}},\nonumber\\
    \alpha_{2} =& ~\frac{r_{-} {\sqrt {\left[ \kappa  n r_{-} r_{+}+i q Q (r_{-}-r_{+})\right]^{2}}}}{r_{+}(r_{+}-r_{-})},
    \nonumber\\
    \alpha_{3} =& ~\frac{\kappa n r_{+}^{2}}{r_{+}-r_{-}},
    \nonumber\\
    a_{1} =& ~l(l+1)-\alpha_{1}-\alpha_{2}-\alpha_{3}-2 \alpha_{3}\left( \alpha _{1}+\alpha _{2} \right)\nonumber\\
    &+\frac{2 \kappa^2 n^2 r_{+}^3 (r_{+}-2 r_{-})}{(r_{+}-r_{-})^2}+\frac{2 i \kappa n q Q r_{+}^2}{r_{+}-r_{-}}+\mu ^2 r_{+}^2,\nonumber\\
    a_{2} =& ~\left(r_{+}^2-r_{-}^2\right) \left(\mu ^2+2 \kappa^2 n^2\right)-2 \alpha_{1} (\alpha_{2}+\alpha_{3}+1)\nonumber\\
    &+\frac{2 i \kappa n q Q \left(r_{+}^2+r_{-} r_{+}-2 r_{-}^2\right)}{r_{+}}\nonumber\\
    &-\frac{2 q^2 Q^2 r_{-} (r_{+}-r_{-})}{r_{+}^2},\nonumber\\
    a_{3} =& ~1+2\alpha_{3},\nonumber\\
    a_{4} =& ~1+2\alpha_{2},\nonumber\\
    a_{5} =&~-2\alpha_{1}.
\end{align}
The confluent Heun function is normalized to unity on the event horizon and we have followed the convention set in {\tt {Mathematica}} for the order of its arguments. 
The expression (\ref{eq:pHeun}) is useful as the confluent Heun function is provided as a built-in function in {\tt {Mathematica}}, and therefore one can generate these mode functions very efficiently. 

The mode solutions $q_{nl}(r)$ can also be written in terms of confluent Heun functions, but the relevant confluent Heun functions in this case are of logarithmic type and not built-in to {\tt {Mathematica}}.
Instead, we generate the $q_{nl}(r)$ by numerically integrating the radial equation (\ref{eq:RNRad}), 
making use of  {\tt {Mathematica}}'s {\tt {NDSolve}} function.
This was achieved through the modification of the numerical integration {\tt {Mathematica}} notebook of the {\tt {ReggeWheeler}} package of the Black Hole Perturbation Toolkit \cite{BHPToolkit}.
We start the integration at a large value of $r$, approximating the solution using an asymptotic expansion, and then integrating inwards with $r$ decreasing.
Since the radial mode functions satisfy (\ref{eq:npm}), we need only compute $p_{nl}(r)$ and $q_{nl}(r)$ for $n\ge 0$.
  
\subsection{Numerical results}
\label{sec:results}

We now use the expressions of the previous section 
(\ref{eq:rencondensate}, \ref{eq:Jtau}, \ref{eq:rset2})
to calculate the renormalized scalar condensate, current and RSET. 
We consider a Reissner-Nordstr\"om black hole with charge $Q=M/2$ and use units in which $M=1$. 
In these units the event horizon is located at $r=r_{+}=1+{\sqrt {3}}/2\approx 1.86602540$ to 9 s.f. (\ref{eq:rplus}).
We assume that the scalar field is minimally coupled to the scalar curvature, so that $\xi =0$, and set the scalar field mass to be $\mu M=1/10$, whilst the scalar field charge is $q M=1/4$.
With these parameters we have $\mu>\frac{qQ}{r_{+}}$ and therefore no classical charge superradiance \cite{DiMenza:2014vpa}. 
In this case the Euclidean Green function (\ref{eq:GE}) is uniquely defined.
The presence of superradiant modes complicates the definition of a Hartle-Hawking-like state for a charged scalar field on a Reissner-Nordstr\"om black hole \cite{Balakumar:2022yvx} and therefore, the construction of the Euclidean Green function.
For this reason, we do not consider the possibility of charge superradiance in this paper, 
although we expect the framework presented in this paper for implementing Hadamard renormalization will be unaffected by the presence of superradiant modes. 
Finally, we also set the arbitrary renormalization length scale $L=M=1$ (\ref{eq:KHad}).

We first generated our radial mode functions $p_{nl}(r)$, $q_{nl}(r)$ using the procedure described in Sec.~\ref{sec:modes}, for $0\le l\le l_{\rm {max}}$ and $0\le n\le n_{\rm {max}}$.
We used a working precision of 100 digits, and found that the Wronskian of $p_{nl}(r)$, $q_{nl}(r)$ was constant across our radial grid to at least 46 digits for all values of $l$ and $n$ used. 

\begin{figure}
    \centering
    \includegraphics[scale=0.85]{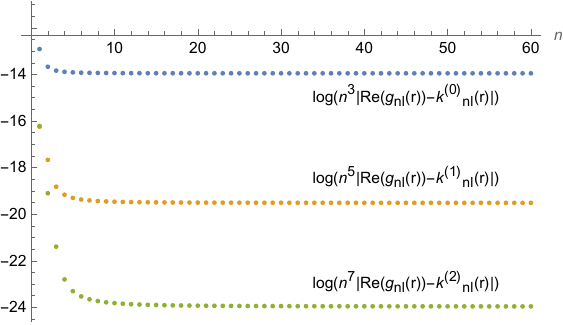}
    \caption{Log plot demonstrating the convergence of the terms in the $n$-sum, for fixed $l$, used to calculate the scalar condensate (\ref{eq:rencondensate}). The blue, yellow and green lines correspond to setting $m=0,1,2$ respectively in the expansion of the relevant part of the Hadamard parametrix (\ref{eq:regmodes}), and show the expected ${\mathcal {O}}(n^{-2m-3})$ behaviour.}
    \label{fig:nPlot}
\end{figure}

\begin{figure}[h]
    \centering
    \includegraphics[scale=0.85]{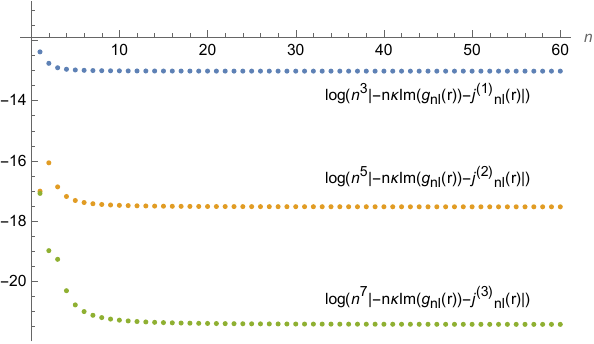}
    \caption{Log plot demonstrating the convergence of the additional terms in the $n$-sum, for fixed $l$, used to calculate the renormalized current (\ref{eq:wtau}). 
    The blue, yellow and green lines correspond to setting $m=1,2,3$ respectively in the expansion of the relevant part of the Hadamard parametrix (\ref{eq:jm}), and show the expected ${\mathcal {O}}(n^{-2m-1})$ behaviour.}
    \label{fig:Current_nPlot}
\end{figure}

The next stage is to compute the mode sums appearing in 
(\ref{eq:rencondensate}, \ref{eq:wtau}, \ref{eq:wtautau}, \ref{eq:wphiphi1}), performing the sum over $n$ first, before the sum over $l$.
As in the neutral case \cite{Taylor:2016edd,Taylor:2017sux}, the summand in the sum relevant for calculating the scalar condensate (\ref{eq:rencondensate}), namely $\{ \Re \left(g_{nl}(r)\right)-k^{(m)}_{nl}(r) \}$, converges as ${\mathcal {O}}(n^{-2m-3})$ for fixed $l$ and large $n$, as demonstrated in 
Fig.~\ref{fig:nPlot} for $m=0,1,2$. 
Similarly, the additional terms relevant for calculating the current (\ref{eq:wtau}), namely $\{-n\kappa \, {\Im }\left(g_{nl}(r)\right)-j_{nl}^{(m)}(r) \} $, converge as ${\mathcal {O}}(n^{-2m-1})$ for fixed $l$ and large $n$, as can be seen in Fig.~\ref{fig:Current_nPlot} for $m=1,2,3$. 
From Fig.~\ref{fig:nPlot}, we deduce that the additional mode sums required for the RSET computation (\ref{eq:wtautau}), namely $\{ n^{2}\kappa ^{2}  [ \Re \left(g_{nl}(r)\right)-k^{(m)}_{nl}(r) ] \} $, also converge as 
${\mathcal {O}}(n^{-2m-1})$ for fixed $l$ and large $n$. Here the real and imaginary parts of $g_{nl}(r)$ arise through the use of the relation (\ref{eq:gnpm}) in the expressions (\ref{eq:GESym}, \ref{eq:GEASym}).

\begin{table*}
\centering
\begin{tabular}{ |p{1cm}|p{3.9cm}|p{3.9cm}|p{3.9cm}|p{3.9cm}|}
\hline
\multicolumn{5}{|c|}{$r/M=1.86645496$}\\
\hline
& $l_{\rm {max}}=30$, $n_{\rm {max}}=30$ & $l_{\rm {max}}=40$, $n_{\rm {max}}=30$ & $l_{\rm {max}}=50$, $n_{\rm {max}}=30$ & $l_{\rm {max}}=60$, $n_{\rm {max}}=30$  \\
\hline
$m=2$ &0.00002618590886922940   &0.00002618502912849888 &0.00002618461774413245 &0.00002618439521262908\\
\hline
$m=3$ &0.00002618382980083290   &0.00002618383654191550 &0.00002618383844230895 &0.00002618383913839172 \\
\hline
$m=4$ &0.00002618383985909550  &0.00002618383981352664 &0.00002618383980593583 &0.00002618383980409387\\
\hline
$m=5$ &0.00002618383980252072   &0.00002618383980307558 &0.00002618383980312958 &0.00002618383980313821 \\
\hline
$m=6$ &0.00002618383980315258 &0.00002618383980314160 &0.00002618383980314098 &0.00002618383980314092 \\

\hline
\end{tabular}
\caption{RSET component $M^{4}\langle\hat{T}^{t}_{~t}\rangle$ to 20 decimal places close to the event horizon, demonstrating its convergence as $m$ and $l_{\text{max}}$ are increased while $n_{\text{max}}$ is held fixed.}
\label{RSETtable:2}
\end{table*}

The number of modes required for final answers for the renormalized expectation values to the desired precision can be reduced by increasing the order $m$ in the expansion of the Hadamard parametrix (\ref{eq:regmodes}, \ref{eq:jm}), at the cost of increased computation time of the required regularization parameters (\ref{eq:Psi}, \ref{eq:chi}). 
Table \ref{RSETtable:2} shows the effect of increasing the expansion order $m$ and $l_{\rm {max}}$ on the convergence of results for the RSET component $\langle\hat{T}^{t}_{~t}\rangle$, with $n_{\rm {max}}$ fixed to be $n_{\rm {max}}=30$. 
Increasing $m$ has the more significant effect on improving convergence than increasing $l_{\rm {max}}$. 
We also found that increasing  $l_{\rm {max}}$ has a greater impact on the rate of convergence than increasing $n_{\rm {max}}$. 
This informed our decision to set $l_{\text{max}}=40$ and $n_{\text{max}}=30$, and, following \cite{Taylor:2022sly}, we set $m=6$.
These choices proved sufficient for yielding an RSET satisfying the conservation equation
\begin{equation}
    \nabla_{\alpha }\langle \hat{T}^{\alpha }{}_{\beta }\rangle = 4\pi F_{\alpha \beta }\langle \hat{J}^{\alpha }\rangle,
\end{equation}
to at least 10 decimal places.

\begin{figure}
    \centering
    \includegraphics[scale=0.65]{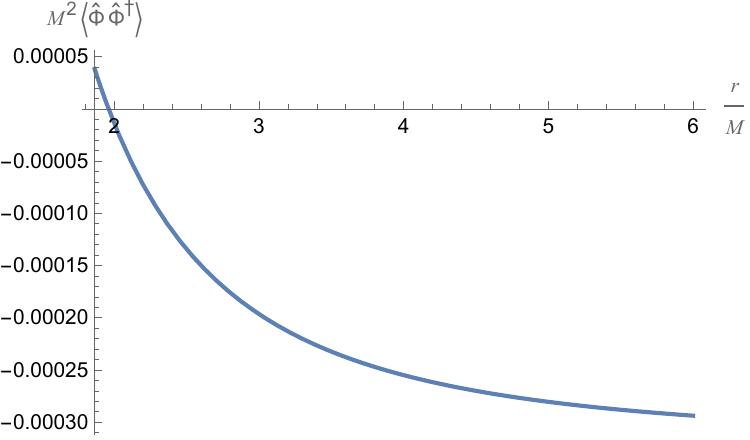}
    \caption{Renormalized scalar condensate $\langle \hat{\Phi}\hat{\Phi}^{\dagger}\rangle $ for a charged scalar field on a Reissner-Nordstr\"om black hole having charge $Q=M/2$. The event horizon radius is $r_{+}/M\approx 1.866$. The scalar field is minimally coupled to the space-time curvature and has mass $\mu M = 1/10$ and charge $qM=1/4$. We use units in which $M=1$.}
    \label{fig:PhiSqrPlot}
\end{figure}

\begin{figure}
    \centering
    \includegraphics[scale=0.65]{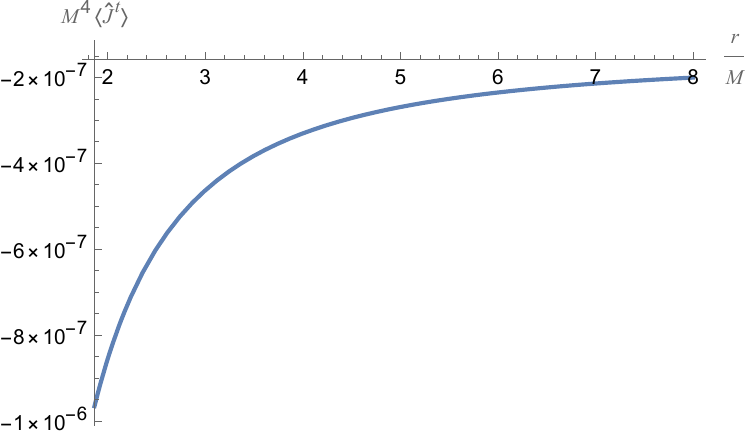}
    \caption{Renormalized time component of the current $\langle {\hat {J}}^{t}\rangle $ for a charged scalar field on a Reissner-Nordstr\"om black hole having charge $Q=M/2$. The event horizon radius is $r_{+}/M\approx 1.866$. The scalar field is minimally coupled to the space-time curvature and has mass $\mu M = 1/10$ and charge $q M=1/4$. We use units in which $M=1$.}
    \label{fig:jtPlot}
\end{figure}

\begin{figure}
    \centering
    \includegraphics[scale=0.65]{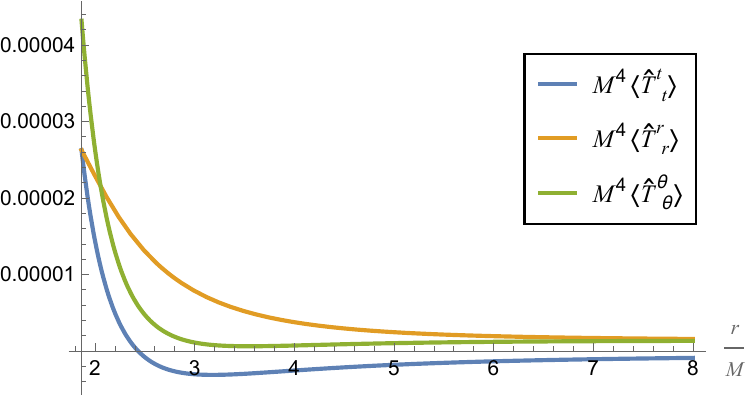}
    \caption{Nonvanishing components of the RSET, $\langle {\hat {T}}^{t}{}_{t}\rangle $, $\langle {\hat {T}}^{r}{}_{r}\rangle $ and $\langle {\hat {T}}^{\theta }{}_{\theta }\rangle =\langle {\hat {T}}^{\phi }{}_{\phi }\rangle $ for a charged scalar field on a Reissner-Nordstr\"om black hole having charge $Q=M/2$. The event horizon radius is $r_{+}/M\approx 1.866$. The scalar field is minimally coupled to the space-time curvature and has mass $\mu M = 1/10$ and charge $q M=1/4$. We use units in which $M=1$.}
    \label{fig:RSETPlot}
\end{figure}

Our numerical results for the scalar condensate $\langle \hat{\Phi}\hat{\Phi}^{\dagger}\rangle $, time component of the current $\langle {\hat {J}}^{t}\rangle $ and the nonzero components of the RSET (namely $\langle {\hat {T}}^{t}{}_{t}\rangle $, $\langle {\hat {T}}^{r}{}_{r}\rangle $ and $\langle {\hat {T}}^{\theta }{}_{\theta }\rangle =\langle {\hat {T}}^{\phi }{}_{\phi }\rangle $)
are presented in Figs.~\ref{fig:PhiSqrPlot}, \ref{fig:jtPlot} and \ref{fig:RSETPlot} respectively.


From Fig.~\ref{fig:PhiSqrPlot}, we see that the scalar condensate is regular everywhere outside the event horizon (see the analysis in Sec.~\ref{sec:horizon} below for its behaviour on the horizon) and monotonically decreasing as the radial coordinate $r$ increases. 
The scalar condensates for massless \cite{Candelas:1984pg} and massive \cite{Anderson:1989vg} scalar fields on a Schwarzschild black hole are similarly monotonically decreasing as $r$ increases.
For a neutral scalar field on a Reissner-Nordstr\"om black hole \cite{Anderson:1990jh}, again the scalar condensate monotonically decreases as $r$ increases, except when the charge of the black hole is close to its maximum value $Q=M$, which is not the situation we are considering here.
We also note that in Fig.~\ref{fig:PhiSqrPlot}, the scalar condensate is negative except for a region close to the event horizon, whereas the scalar condensate is positive everywhere outside the event horizon for the neutral scalar field results in Refs.~\cite{Candelas:1984pg,Anderson:1989vg,Anderson:1990jh} (however, this is not the case for a neutral scalar field on a higher-dimensional black hole \cite{Taylor:2016edd,Taylor:2017sux}). 

The only nonzero component of the renormalized current is $\langle {\hat {J}}^{t}\rangle $, shown in Fig.~\ref{fig:jtPlot}. 
We find that this is negative everywhere outside the event horizon, and monotonically increasing as the radial coordinate $r$ increases. 
Comparing the results in Fig.~\ref{fig:jtPlot} with those in Figs.~\ref{fig:PhiSqrPlot} and \ref{fig:RSETPlot}, it can be seen that the renormalized current has a magnitude which is about 100 times smaller than the magnitudes of the scalar condensate or RSET components.
The magnitude of $\langle {\hat {J}}^{t}\rangle $ depicted in Fig.~\ref{fig:jtPlot} is of similar order of magnitude to that computed for a charged scalar on a Reissner-Nordstr\"om-de Sitter black hole in Ref.~\cite{Klein:2021les}.

The RSET has three nonzero components,  $\langle {\hat {T}}^{t}{}_{t}\rangle $, $\langle {\hat {T}}^{r}{}_{r}\rangle $ and $\langle {\hat {T}}^{\theta }{}_{\theta }\rangle =\langle {\hat {T}}^{\phi }{}_{\phi }\rangle $, and these can be seen in  Fig.~\ref{fig:RSETPlot}.
As with the scalar condensate and current, all three components are regular everywhere outside the event horizon.
Once again, the results depicted in Fig.~\ref{fig:RSETPlot} share some qualitative features with the corresponding RSET components for a neutral scalar field on a Schwarzschild  \cite{Howard:1984qp,Howard:1984ttx} (at least when the neutral scalar field is conformally coupled \cite{Taylor:2022sly})  and on  a Reissner-Nordstr\"om black hole \cite{Anderson:1993if,Anderson:1994hg} (again when the field is conformally coupled and providing the black hole charge is not too large). 
The component $\langle {\hat {T}}^{t}{}_{t}\rangle $ is positive on the event horizon and monotonically decreasing as $r$ increases close to the horizon.  
This component has a minimum at $r\sim 3$, and is monotonically increasing for larger $r$, although it remains negative for all $r$ larger than the location of the minimum.
In contrast, the remaining components, $\langle {\hat {T}}^{r}{}_{r}\rangle $  and
$\langle {\hat {T}}^{\theta }{}_{\theta }\rangle $, are positive everywhere outside the horizon.
The  radial component $\langle {\hat {T}}^{r}{}_{r}\rangle $ is monotonically decreasing as $r$ increases, while the angular component $\langle {\hat {T}}^{\theta }{}_{\theta }\rangle $ is decreasing close to the horizon, has a minimum at $r\sim 3$ and is increasing for larger values of $r$.
Far from the black hole, the spatial components $\langle {\hat {T}}^{r}{}_{r}\rangle $ and $\langle {\hat {T}}^{\theta }{}_{\theta }\rangle$ converge towards a common value, while
on the horizon, the interpolated values of $\langle \hat{T}^{r}_{~~r}\rangle$ and $\langle \hat{T}^{t}_{~~t}\rangle$ at the horizon agree to 17 decimal places. 
Next, we further explore the behaviour of all the renormalized expectation values on the horizon.

\subsection{Expectation values on the event horizon}
\label{sec:horizon}
In this section we outline the calculation of the scalar condensate, renormalized current and RSET components on the event horizon. 
We will use these calculated values as a check on the numerical off-horizon results presented in the previous section. 

To obtain the horizon values, we exploit the choice of gauge in (\ref{eq:gauge}), which has the effect that the radial equation (\ref{eq:Rad}) reduces to its neutral equivalent as the horizon is approached. This in turn means that the solutions $p_{nl}(r)$ and $q_{nl}(r)$ have the following leading order behaviour at the horizon:
\begin{equation}
    p_{nl}(r)\sim(r-r_+)^{|n|/2}, \qquad q_{nl}(r)\sim(r-r_+)^{-|n|/2}.
\end{equation}
Therefore choosing to point-split purely in the radial direction and placing the innermost point, the argument of $ p_{nl}(r)$, on the horizon will collapse the sum over $n$ in the various mode sums required. For example, the mode sums required  for the scalar condensate and the current, (\ref{eq:GESym}) and (\ref{eq:Gasymtau}) respectively, reduce to
\begin{align}
\label{eq:Gh}
  \{G^{\sy}(x,x')\} & = \frac{\kappa }{8\pi^2} \sum_{l=0}^{\infty}(2l+1)\mathcal{N}_{0l} q_{0l}(r)
  \nonumber \\
     \{\partial_{\tau}G^{\asy}(x,x')\} & =0,
\end{align}
where the notation $\{ \, \}$ denotes that the partial coincidence limit $\tau'\to \tau $, $\theta'\to\theta $, $\phi'\to\phi$ has been taken.
This procedure also yields:
\begin{multline}
\label{eq:gpph}
    \{\nabla_{\phi}\nabla^{\phi'}G^{\sy}(x,x')\}
    \\
    =\frac{\kappa }{8\pi^{2}r_+}\sum_{l=0}^{\infty}\frac{l}{r}(l+1)(2l+1) \mathcal{N}_{0l} q_{0l}(r).
\end{multline}
For a neutral massless field, the solution $q_{0l}(r)$ can be expressed in terms of Legendre functions and the above modes sums can be computed exactly \cite{Candelas:1980}. However for a massive field, this is no longer the case and we need an alternative approach. 

One such approach is that of \cite{Breen:2010ux,Breen:2011af}, where uniform approximations for the radial solution $q_{nl}(r)$ were used to calculate the event horizon values of the scalar condensate and the renormalized stress energy tensor for a field in the Hartle-Hawking state. Even though this method was developed for the case of a neutral scalar field, the results for any mode sums involving the $n=0$ mode $q_{0l}(r)$, such as (\ref{eq:Gh}, \ref{eq:gpph}), carry through to the charged case with only minor adjustments, leading to the following quasi-analytical expression for the scalar condensate on the event horizon:
\begin{align}
\label{eq:phihor}
  &\langle \hat{\Phi}\hat{\Phi}^{\dagger}\rangle_{\textrm{ren}}= w=\frac{1}{8\pi^2 r_+^2}\bigg\{ \frac{1}{12} +
\frac{d}{dx} \zeta\left(x,\textstyle{\frac{1}{2}} 
+ i\delta\right)\bigg|_{x=-1} \nonumber\\
& +\frac{d}{dx} \zeta\left(x,\textstyle{\frac{1}{2}} - i\delta\right)\bigg|_{x=-1}-i \delta\ln\left[\frac{\Gamma\left(\textstyle{\frac{1}{2}} + i\delta\right)}{\Gamma\left(\textstyle{\frac{1}{2}} - i\delta\right)}\right]\nonumber\\
& + \mu^2r_+^2\left[1  +\gamma-\ln\left(\frac{2r_+}{L} \right)\right]+\frac{\kappa r_+}{3} +\sum^{\infty}_{l=0}(2l+1)\beta_l \bigg\} ,
\end{align}
where  $\delta^2=\mu^2 r_+^2+1/12$ is a constant,  $\zeta$  is the generalized Riemann Zeta function, $\gamma$ is Euler's constant and $\Gamma$ is Euler's gamma function. The quantity $\beta_l$ ensures that the approximation to $q_{0l}(r)$ employed contains the appropriate multiples of $p_{0l}(r)$ and must be obtained numerically. The details of how these $\beta_l$ terms are obtained are given in App.~\ref{sec:ApA}.  We note that while the charge of the scalar field, $q$, does not appear explicitly in the above expression for $w$, the quantity $w$ does in fact depend on $q$ through the numerical $\beta_l$ terms.

\begin{figure}
    \centering
    \includegraphics[scale=0.65]{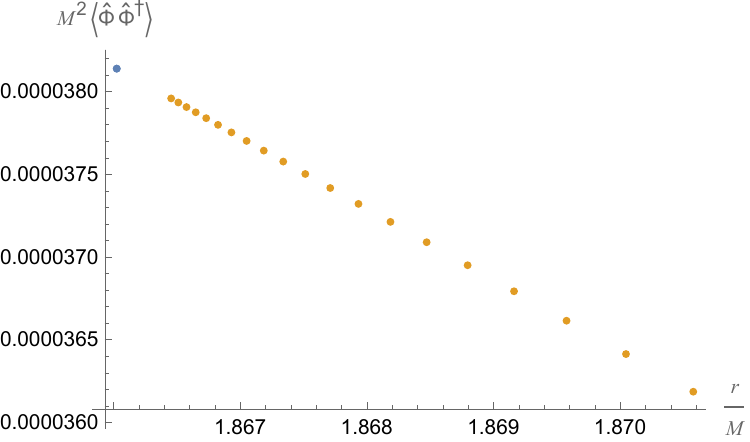}
    \caption{Renormalized scalar condensate $\langle \hat{\Phi}\hat{\Phi}^{\dagger}\rangle $ near the black hole event horizon. Yellow dots are computed numerically using the method in Sec.~\ref{sec:results}, while the blue dot is the on-horizon result (\ref{eq:phihor}).}
    \label{fig:phi2hor}
\end{figure}

In Fig.~\ref{fig:phi2hor} we compare the results of our numerical computations from the previous section (yellow dots) with the on-horizon expression (\ref{eq:phihor}), shown as a blue dot. The value obtained by extrapolating the off-horizon results to the horizon agrees with the quasi-analytical expression (\ref{eq:phihor}) to nine significant figures.

\begin{figure}
    \centering
    \includegraphics[scale=0.65]{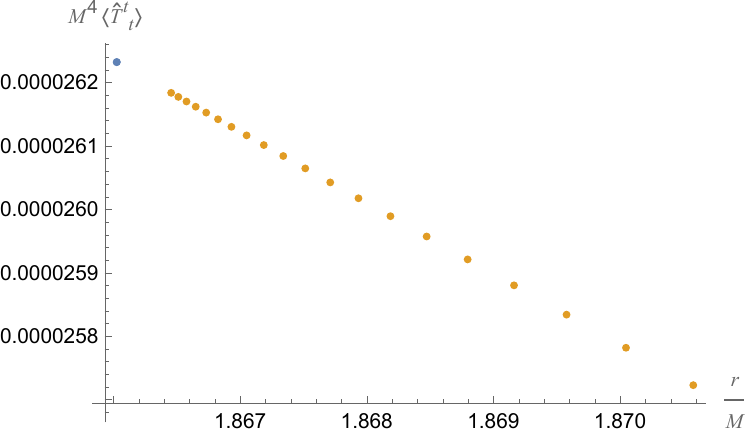}
    \\
    \includegraphics[scale=0.65]{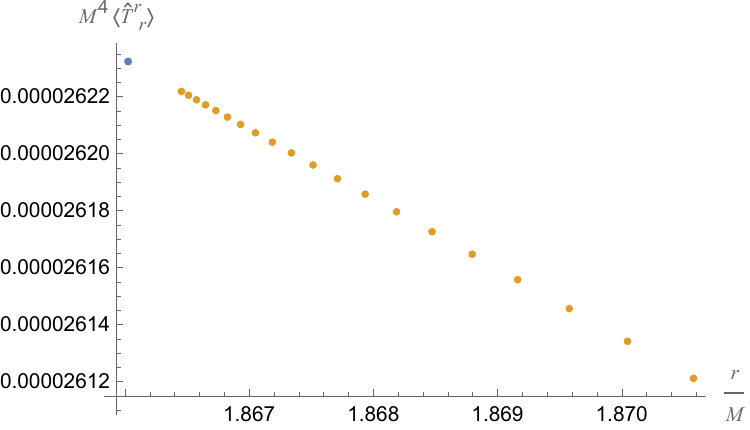}
    \\
    \includegraphics[scale=0.65]{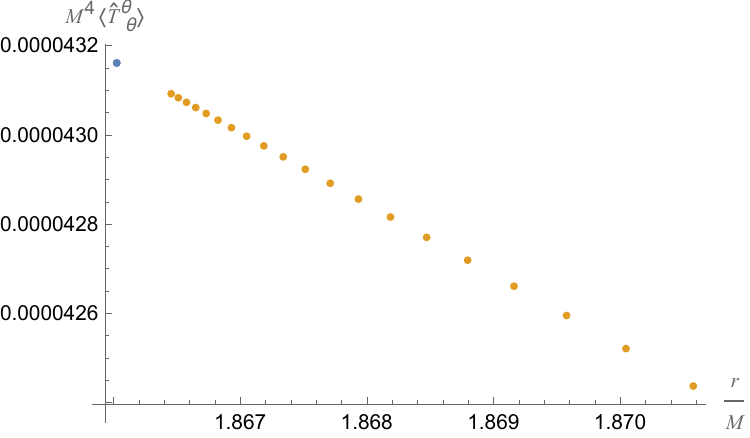}
    \caption{Nonvanishing RSET components $\langle {\hat {T}}^{t}{}_{t}\rangle $ (top), $\langle {\hat {T}}^{r}{}_{r}\rangle $ (middle) and $\langle {\hat {T}}^{\theta }{}_{\theta }\rangle $ (bottom) near the black hole event horizon. Yellow dots are computed numerically using the method in Sec.~\ref{sec:results}, while blue dots are the on-horizon results (\ref{eq:Trrhor}, \ref{eq:Tpphor}).}
    \label{fig:SEThor}
\end{figure}

Employing a similar approach to the mode sum in (\ref{eq:gpph}) leads to a similar, if much longer, quasi-analytical expression for $\widetilde{w}^{\phi}{}_{\phi}$ on the horizon, which is also presented in App.~\ref{sec:ApA}. 
Similarly, at $r=r_{+}$ both $\widetilde{w}^{t}{}_{t}$ and $ \widetilde{w}^{r}{}_{r}$  can be expressed terms of the $n=0$ mode $q_{0l}(r)$, and hence we have, from the results of \cite{Breen:2011af}, that on the event horizon
\begin{equation}
  \widetilde{w}^{t}{}_{t}= \widetilde{w}^{r}{}_{r}=-\widetilde{w}^{\phi}{}_{\phi}+\frac{\mu^2}{2} w-\frac{3v_
1}{8\pi^2}.
\end{equation}
 Therefore, using (\ref{eq:rset2}) and noting that $A_{\tau}$ vanishes on the event horizon, we see that all components of the renormalized stress energy tensor on the horizon may be expressed in terms of $\widetilde{w}^{\phi}{}_{\phi}$, $w$ and derivatives of $w$. In fact,
for minimal coupling we see that
\begin{equation}
    \langle \hat{T}^{r}{}_{r}\rangle =   \langle \hat{T}^{t}{}_{t}\rangle=\widetilde{w}^{\phi}{}_{\phi}-\frac{\mu^2}{2}w +\frac{v_
1}{4\pi^2}.
\label{eq:Trrhor}
\end{equation}
Inserting Eqs. (\ref{eq:phihor}) and (\ref{eq:wphiphi}) into the above expression  and evaluating for the parameter set considered in this paper, gives 
the result shown in Fig.~\ref{fig:SEThor}. We find agreement with the off-horizon numerical results extrapolated to the horizon to nine significant figures.

For the final stress energy tensor component, we have, with $\xi=0$, that on the horizon:
\begin{equation}
    \langle \hat{T}^{\phi}{}_{\phi}\rangle =  -\widetilde{w}^{\phi}{}_{\phi}-\frac{\kappa}{2}w_{,r} -\frac{v_1}{8\pi^2}.
    \label{eq:Tpphor}
\end{equation}
To calculate $w_{,r}$ directly requires taking derivatives at both $r$ and $r'$, with the derivative at $r'$ leading to a mode sum involving the $n=1$ mode. In this case, the radial equation is now complex and the neutral results do not carry over as easily as for the $n=0$ case. However, we may circumvent this issue by exploiting the fact that we have both the exact value of $w$ on the horizon and extremely accurate numerical results in the near-horizon region. We can therefore generate an interpolating function for $w$ that is valid on the horizon, which (numerically) gives its derivative there. 
Taking this approach yields a value shown in Fig.~\ref{fig:SEThor}. We find that this  agrees with the off-horizon results to the same accuracy as for the other RSET components.   

\begin{figure}
    \centering
    \includegraphics[scale=0.65]{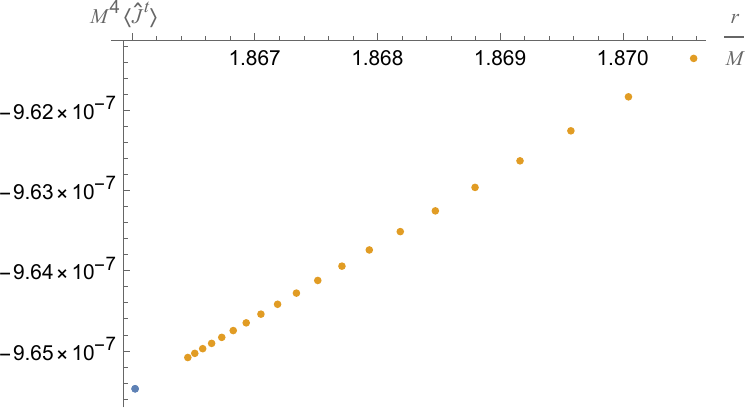}
    \caption{Renormalized time component of the current $\langle {\hat {J}}^{t}\rangle $ near the black hole event horizon. Yellow dots are computed numerically using the method in Sec.~\ref{sec:results}, while the blue dot is the on-horizon result (\ref{eq:Jhor}).}
    \label{fig:Jhor}
\end{figure}

Finally, we turn our attention to the renormalized current. For radial separation, with the innermost point on the horizon, one can show that like $\{\partial_{\tau}G^{\asy}(x,x')\}$, the partial coincidence limit $\{\partial_{\tau} K(x,x')\}$ also vanishes and hence $w_{\tau}$ is zero on the event horizon. Recalling the expression (\ref{eq:Jtau}) for the only nonzero component of the current, 
and noting once more that $A_{\tau}$ vanishes on the horizon, we obtain the result that $\langle \hat{J}_{t}\rangle_{\textrm{ren}}=0$ on the event horizon, which is in agreement with the numerical results. 
To confirm the near-horizon results for $\langle \hat{J}^{t}\rangle_{\textrm{ren}}$ as plotted in Fig. 
 \ref{fig:jtPlot}, we must consider the ratio $\langle\hat{J}_{\tau}\rangle_{\textrm{ren}}/f(r)$ as $r \to r_+$. As both quantities vanish in this limit, we have that 
\begin{equation}
\langle\hat{J}^{t}\rangle_{\textrm{ren}}|_{r=r_+} =\frac{i}{2 \kappa}\frac{d}{d r}\langle\hat{J}_{\tau}\rangle_{\textrm{ren}} \bigg |_{r=r_+}.
\label{eq:Jhor}
\end{equation}
As for the scalar condensate, we may calculate the derivative of the current by exploiting our knowledge of the exact horizon value to obtain the derivative there numerically, without recourse to evaluating any sums involving the $n=1$ mode. The result of this computation is shown in Fig.~\ref{fig:Jhor}. We find agreement with the extrapolated horizon values to seven significant figures.

\section{Conclusions}
\label{sec:conc}

We have presented a new, practical, method for the computation of renormalized expectation values for a charged quantum scalar field.
Our scheme is applicable to any static, spherically-symmetric space-time with a time-independent background electrostatic potential.
Our methodology was developed from the ``extended coordinates'' approach for a neutral scalar field \cite{Taylor:2016edd,Taylor:2017sux,Taylor:2022sly}.
Working on a Euclidean space-time, we first constructed the point-split Euclidean Green function as a mode sum over separable solutions of the charged scalar field equation, assuming that the field is in a thermal Euclidean state.
The Euclidean Green function is renormalized by subtracting the Hadamard parametrix \cite{Balakumar:2019djw}.
We first wrote the Hadamard parametrix in terms of ``extended coordinates'', from which it can be expressed as a mode sum, enabling the renormalization to be performed mode-by-mode.
Using the renormalized Green function, we derived expressions for the renormalized scalar condensate, current and stress-energy tensor.
We have demonstrated the effectiveness of our method by applying it to a particular case of a massive, minimally-coupled, charged scalar field in the Hartle-Hawking state on a Reissner-Nordstr\"om black hole.

Our method is sufficiently efficient that it would now be practical to explore the dependence of the renormalized expectation values on the scalar field mass, charge, and coupling to the Ricci scalar curvature.
Previous work for neutral scalar fields on both Schwarzschild \cite{Taylor:2022sly} and Reissner-Nordstr\"om \cite{Anderson:1993if,Anderson:1994hg,Arrechea:2023fas} black holes have shown that expectation values depend strongly on the scalar field parameters.
In addition, our results depend on a renormalization length scale $L$. Changing this scale may also affect the qualitative features of the renormalized expectation values. 

On black hole space-times, our approach most naturally applies to the Hartle-Hawking state (although the ``extended coordinates'' method has very recently been generalized to the Boulware state \cite{Arrechea:2024cnv}).
However, differences in expectation values between two quantum states do not require renormalization and are therefore easier to compute.
Our results can therefore also be extended to different quantum states, such as the Unruh \cite{Unruh:1976db} and Boulware \cite{Boulware:1974dm} states  (see \cite{Arrechea:2023fas} for
recent results on expectation values in these states for a neutral scalar field on a Reissner-Nordstr\"om black hole). 
While we have focussed on a particular Reissner-Nordstr\"om black hole, it would also be feasible to study the effect of changing the charge of the black hole (on which the expectation values for a neutral scalar field also depend strongly \cite{Anderson:1993if,Anderson:1994hg,Arrechea:2023fas}).
Furthermore, our implementation works on any static, spherically symmetric, space-time, and hence is applicable to charged quantum scalar fields on more general black hole space-times, such as Reissner-Nordstr\"om-(anti-)de Sitter. 
In this paper, we have worked in four space-time dimensions, although the ``extended coordinates'' method works equally well in dimensions greater than four (at least for the scalar condensate) \cite{Taylor:2016edd,Taylor:2017sux}, so we anticipate that our results in this paper could also be extended to higher dimensions.

Finally, in this paper we have restricted our attention to the region exterior to the event horizon of a black hole.
There has been a great deal of recent interest in quantum effects on black hole interiors, particularly in the behaviour of the RSET near the inner horizon and the consequences for cosmic censorship \cite{Hollands:2019whz,Zilberman:2019buh,Hollands:2020qpe,Juarez-Aubry:2021tae,Klein:2021ctt,Zilberman:2022aum,McMaken:2022dqc,Hintz:2023pak,Klein:2023urp,McMaken:2023uue,Klein:2024sdd,McMaken:2024fvq,Zilberman:2024jns}.
With the notable exception of \cite{Klein:2021ctt}, work to date on this topic has largely focused on a neutral scalar field, even when the background black hole is charged.
Given that the analysis of the current on the black hole interior has revealed unexpected (dis)charge processes \cite{Klein:2021ctt}, it would be of great interest to study the RSET for a charged scalar near the inner horizon of a charged black hole.

\begin{acknowledgments}
The authors wish to thank Eoin Scanlon for helpful conversations at the early stages of this project. The work of E.W.~is supported by STFC grant number ST/X000621/1.
E.W.~also acknowledges support of the Institut Henri Poincar\'{e} (UAR 839 CNRS-Sorbonne Universit\'{e}), and LabEx CARMIN (ANR-10-LABX-59-01).
This work makes use of the Black Hole Perturbation Toolkit \cite{BHPToolkit}.
G.M. and E.W. acknowledge IT Services at The University of Sheffield for the provision of services for High Performance Computing.
Data supporting this publication can be freely downloaded from the University of Sheffield research data repository at \cite{data}, under the terms of the Creative Commons Attribution (CC-BY) licence.
\end{acknowledgments}

\appendix

\section{Coefficients in the Hadamard expansion}
\label{app:hadamardcoeff}
In this appendix, we give explicitly the coefficients of the Hadamard parametrix when expanded in ``extended coordinates'' (with $\Delta r=0$), that is, the coefficients appearing in Eqs.~(\ref{eq:HadamardSym}) and (\ref{eq:HadamardASym}). The assumptions are that the space-time is a static, spherically symmetric black hole space-time with surface gravity $\kappa$ and that there is a background electrostatic potential. Otherwise, the metric function $f(r)$ and the Euclideanized potential $A_{\tau}$ are arbitrary functions of $r$. We list the coefficients required to expand the parametrix up to the order such that  $\mathcal{O}(\epsilon^{4}\log\,\epsilon)$ terms are ignored, with $\Delta x\sim\epsilon$. Subtracting the parametrix up to this order is sufficient to regularize the stress-energy tensor for a massive charged scalar in an arbitrary static spherically, symmetric space-time. In practice, we subtract a much higher-order expansion of the parametrix which serves to accelerate the convergence of the mode-sum representation of the renormalized stress-energy tensor. The higher order coefficients are too lengthy to be of use in print form, but we include in the Supplemental Material a {\tt {Mathematica}} Notebook with these expressions specialised to Ricci flat space-times and imposing the Einstein-Maxwell equations on the vector potential. Below, all derivatives ($f'$, $f^{(n)}$ etc.) are with respect to the radius $r$. We omit the explicit dependence on $r$ throughout for typographical convenience.
\begin{widetext}
\begin{align}
    \mathcal{DE}^{(\supR)}_{00}= & ~ 2,\nonumber\\
    \mathcal{DE}^{(\supR)}_{10}= & ~ -q^2 A_{\tau}^2-\frac{f\, \left(r^2 f''-2 r\, f'+2 f-2\right)}{12 \,r^2},\nonumber\\
    \mathcal{DE}^{(\supR)}_{11}= & ~ \frac{f\, \left(r^2 \left[f'^2-4 \kappa ^2\right]-4 f\, \left[r\, f'+1\right]+4
   f^2\right)}{24 \,r^2},\nonumber\\
    \mathcal{DE}^{(\supR)}_{20}= & ~ \frac{1}{2880\, r^4}\Bigg(240 q^2 r^3  \,f \, A_{\tau} \, A_{\tau}'\left[r \,f'-2
   f\right]-120 q^2 r^2 A_{\tau}^2 \Big\{ f \left[-r^2
   f''+2 r \, f'+2\right]-2 f^2+2 \kappa ^2 r^2\Big\} +240 q^4 r^4
   A_{\tau}^4\nonumber\\
   & +f \Big\{ -5 r^2 \left[4 \kappa ^2-f'^2\right] \left[r^2
   f''-2 r \,f'-2\right]-8 f^2 \big[3 r^3 f^{(3)}-7 r^2 f''+19 r\,
   f'+10\big]\nonumber\\
   &+f \left[ 9 r^4 f''^2-20 r^2 f''+86 r^2 f'^2+4 r\, f'
   \big(3 r^3 f^{(3)}-14 r^2 f''+20\big)-40 \kappa ^2 r^2+4\right]+76
   f^3\Big\} \Bigg),\nonumber\\
   \mathcal{DE}^{(\supR)}_{21}= & ~\frac{f}{2880 \,r^4}\Bigg(60 q^2 r^2 A_{\tau}^2 \Big\{ r^2 \left[4 \kappa
   ^2-f'^2\right]+4 f\, \left[r\, f'+1\right]-4 f^2\Big\} -r^4 \left[-20
   \kappa ^2 f'^2+f'^4+64 \kappa ^4\right] \nonumber\\
   &+r^2 f\, \Big\{ 20 \kappa ^2
   \left[r^2 f''-6\right]-120 \kappa ^2 r \,f'+30 r \,f'^3+f'^2 \left[30-11 r^2
   f''\right]\Big\}+f^3 \left[-44\, r^2 f''+208\, r \,f'+160\right]\nonumber\\
   &+2 f^2
   \Big\{ 10 r^2 f''-67 r^2 f'^2+f' \left[22 r^3 f''-80 r\right]+60 \kappa ^2
   r^2-28\Big\}-104 f^4\Bigg),\nonumber\\
   \mathcal{DE}^{(\supR)}_{22}= & ~\frac{f^2\, \Big(r^2 \left[f'^2-4 \kappa ^2\right]-4 f\, \left[r\, f'+1\right]+4
   f^2\Big)^2}{1152\, r^4},\nonumber\\
   \mathcal{DE}^{(\supP)}_{11}= & ~-\frac{f'}{6\, r},\nonumber\\
   \mathcal{DE}^{(\supP)}_{21}= & ~ \frac{120 q^2 r^2 \, f\, A_{\tau}\, A_{\tau}'+60 q^2 r^2 A_{\tau}^2\, f'+f\, \Big(-9 r\, f'^2+6 r \,f \big[r \,f^{(3)}-2 f''\big]+2
   f'\, \left[7 r^2 f''+f+5\right]\Big)}{720 \,r^3},\nonumber\\
   \mathcal{DE}^{(\supP)}_{22}= & ~ \frac{7 r^2 f'^2-10 r\, f'+r\, f \,\left(9 r \,f''+4 \,f'\right)-3 f^2+3}{720\, r^4},\nonumber\\
   \mathcal{TE}^{(\supR)}_{10}= & ~\frac{f}{576 \,r^4} \Big(r^2 \left[4 \kappa ^2-f'^2\right]+4 f\, \left[r\, f'+1\right]-4
   f^2\Big)  \nonumber\\ &  \qquad \times \Big(-6 \,\xi\,  r^2 f''+r^2 f''
   +4 [1-6 \xi ] r\, f'+2[1-6\, \xi ]
   f+12\, \xi +6 \mu ^2 r^2-2\Big),\nonumber\\
   \mathcal{TE}^{(\supP)}_{10}= & ~\frac{\Big(f-1\Big) \Big(6\, \xi\,  r^2 f''-r^2 f''+4\, [6 \,\xi -1] r\, f'+2 [6\, \xi -1]\,
   f-12\, \xi -6 \mu ^2 r^2+2\Big)}{144 \,r^4},\nonumber\\
   \mathcal{TE}^{(\supP)}_{11}= & ~-\frac{f\, \Big(r \,f'-2 f+2\Big) \Big(-6\, \xi  r^2 f''+r^2 f''+4 [1-6\, \xi ]
   r\, f'+2[1-6\, \xi ]\, f+12\, \xi +6 \mu ^2 r^2-2\Big)}{144\, r^4},\nonumber\\
   \mathcal{TE}^{(\supL)}_{00}= & ~\frac{-6\, \xi \, r^2 f''+r^2 f''+4 (1-6\, \xi )\, r\, f'+2(1-6\, \xi )\, f+12\, \xi +6 \mu ^2
   r^2-2}{12\, r^2},\nonumber\\
   \mathcal{TE}^{(\supL)}_{10}= & ~\frac{1}{480 \,r^4}\Bigg(-10 q^2 r^4 A_{\tau}'^2-60 \mu ^2 \xi  \,r^4 f''+10 \mu ^2 r^4 f''+30\,
   \xi ^2 r^4 f''^2-10\, \xi\,  r^4 f''^2+r^4 f''^2-120\, \xi^2 r^2 f''\nonumber\\
   &+20\, \xi\,  r^2
   f''-240 \mu ^2 \xi\,  r^3 f'+20 \mu ^2 r^3 f'+480\, \xi^2 r^2 f'^2-100\, \xi\, 
   r^2 f'^2+4 r^2 f'^2-480\, \xi^2 r\, f'+80\, \xi \, r \,f'\nonumber\\
   &+10\, \xi\,  r^4 f^{(3)}
   f'-2 r^4 f^{(3)} f'+240\, \xi^2 r^3 f' \,f''-20\, \xi\,  r^3 f'\, f''-4 r^3
   f'\, f''+2 f\, \Big\{ 5\, \xi\,  r^4 f^{(4)}-r^4 f^{(4)}+40\, \xi\,  r^3 f^{(3)}\nonumber\\
   &-7
   r^3 f^{(3)}+4 \left[15\, \xi^2+10 \,\xi -2\right] r^2 f''+2 \left[120 \,\xi^2-40\, \xi
   +3\right]\, r\, f'-120\, \xi^2+20 \,\xi -60 \mu ^2 \xi\,  r^2\Big\}\nonumber\\
   &+4 \left[30\, \xi^2-10\, \xi
   +1\right]\, f^2+120\, \xi^2+30 \mu ^4 r^4+120 \mu ^2 \xi \, r^2-4\Bigg),\nonumber\\
   \mathcal{TE}^{(\supL)}_{11}= & ~\frac{1}{480 \,r^4}\Bigg(f\, \bigg\{ 20 q^2 r^4 A_{\tau}'^2-10 \mu ^2 r^4 f''+10\, \xi\,  r^4
   f''^2-2 r^4 f''^2-20\, \xi\,  r^2 f'^2+2 r^2 f'^2+120\, \xi\,  r\, f'\nonumber\\
   &-20 r\,
   f'+10\, \xi\,  r^4 f^{(3)} f'-r^4 f^{(3)} \,f'+80\, \xi\,  r^3 f'\, f''-12 r^3
   f'\, f''+r \,f\, \Big((16-80\, \xi ) \,f'\nonumber\\
   &+r\, \big[ r^2 f^{(4)}+(6-20 \,\xi )\, r
  \, f^{(3)}+(12-80\, \xi )\, f''+20 \mu ^2\big]\Big)+8 (5\, \xi -1) \,f^2-40\, \xi -20
   \mu ^2 r^2+8\bigg\} \nonumber\\
   &+40 q^2 r^3 A_{\tau}\, f\, \left(r\, A_{\tau}''+2
   A_{\tau}'\right)-20 q^2 r^2 A_{\tau}^2 \Big\{ -6\, \xi\,  r^2 f''+r^2
   f''+4 (1-6\, \xi )\, r \,f'\nonumber\\
   &+(2-12\, \xi )\, f+12\, \xi +6 \mu ^2 r^2-2\Big\} \Bigg),\nonumber\\
   \mathcal{DO}^{(\supR)}_{00}= & ~ 2 i q A_{\tau},\nonumber\\
    \mathcal{DO}^{(\supR)}_{10}= & ~ -\frac{i q \Big(r\, f \,A_{\tau}'\, \big[r \,f'-2 f\big]+A_{\tau}
   \Big\{ f\, \left[r^2 f''-2 r\, f'-2\right]+2 f^2-\kappa^2
   r^2\Big\} +4 q^2 r^2 A_{\tau}^3\Big)}{12\ r^2},\nonumber\\
    \mathcal{DO}^{(\supR)}_{11}= & ~\frac{i q\, A_{\tau}\, f \Big(r^2 \left[f'^2-4 \kappa ^2\right]-4 f
   \left[r \,f'+1\right]+4 f^2\Big)}{24 \,r^2},\nonumber\\
   \mathcal{DO}^{(\supP)}_{11}= & ~ -\frac{i q \left(f\, A_{\tau}'+A_{\tau}\, f'\right)}{6\, r},\nonumber\\
      \mathcal{TO}^{(\supL)}_{00}= & ~\frac{i q \Big(A_{\tau} \Big\{ -6 \xi  r^2 f''+r^2 f''+4 [1-6 \xi ] r\,
   f'+2[1-6 \xi ] f+12 \xi +6 \mu ^2 r^2-2\Big\}-r\, f \left[r\, A_{\tau}''+2 A_{\tau}'\right]\Big)}{12\, r^2}.
\end{align}

\section{Details of the on-horizon calculations}
\label{sec:ApA}

In this appendix we present an expression for $\hat{w}^{\phi}{}_{\phi}$  on the event horizon and provide further details on the calculation of the $\beta_l$ terms in (\ref{eq:phihor}). 

Taking the approach outlined in Sec.~\ref{sec:horizon} leads to the following expression for $\widetilde{w}^{\phi}{}_{\phi}$  on the event horizon:
\begin{align}
\label{eq:wphiphi}
\widetilde{w}^{\phi}{}_{\phi}  = &  ~ \frac{1}{8\pi^2 r_+^4}\left\{ 
2  \alpha  -\frac{1}{240 r^2_+} \left[ 60 r_{-}^2-5 \left(12 \delta ^2+23\right) r_{-} r_{+}+\left\{ (45-60 \gamma ) \delta
   ^4+(75-30 \gamma ) \delta ^2+53\right\}  r_{+}^2\right]
\right. \nonumber \\ & \left. 
+\frac{1}{360r_+^2}\bigg[ -83
   r_{-}^2
  -15 \left(q^2 Q^2-3 \mu ^4
   r_{+}^4+4 \mu ^2 r_{+}^2+6\right)  r_{+}^2 
   +4 \left(15
   \mu ^2 r_{+}^2+44\right) r_{-} r_{+} 
  \right. \nonumber \\ & \left. 
  + 15 \left(q^2 Q^2+2 \mu ^2 r_{-}
   r_{+}+3 \mu ^4 r_{+}^4\right) r_{+}^2 \log \left(\frac{ L^2}{r_+^2 }\right)\bigg]
  \right. \nonumber\\ 
 &   \left. -\frac{1}{4}\sum_{l=0}^{\infty} \left[
 l(l+1)(2l+1)\left\{ \ln(\nu_l) -\psi(\tfrac{1}{2}+\nu_l)\right\} +\frac{4 \alpha}{3(l+1)} \right] 
 -\int^{\infty}_{0} d\lambda \,  \frac{ \lambda(\lambda^2 +\tfrac{1}{4})}{1 +e^{2\pi \lambda}}\log\left(|\delta^2-\lambda^2| \right)
 \right. \nonumber \\ & \left. 
 -\frac{1}{2}\sum_{l=0}^{\infty}l(l+1)(2l+1)\widetilde {\beta}_{l}\right\},
\end{align}
where $r_{\pm}$ are given in  (\ref{eq:rplus}), $\psi $ is the digamma function, 
\begin{equation}
    \alpha =\frac{1}{4} \mu^2 r^3_{+}\kappa-\frac{Q^2
   q^2}{16}-\frac{1}{240},
   \qquad \delta^2=\mu^2 r_+^2+\frac{1}{12},
   \qquad \nu_{l}=\frac{\mu^2 r_+^2+1/3 +l(l+1)}{8\sqrt{|\alpha|}},
 \end{equation}
 and $\gamma$ is Euler's constant.
 \end{widetext}
  The quantity $\widetilde {\beta}_l$ ensures that the approximation to $q_{0l}(r)$ employed contains the appropriate multiple of $p_{0l}(r)$ and must be obtained numerically.
 The sums and integrals in the final line of (\ref{eq:wphiphi}) must also be computed numerically. 
 
 Both $\widetilde {\beta}_l$ in the above expression  (\ref{eq:wphiphi}) and the corresponding $\beta_{l}$ term in the expression for the scalar condensate in (\ref{eq:phihor}) arise by expressing the full $q_{0l}(r)$ in the following way:
 \begin{align}
 \label{eq:approx}
     q_{0l}(r)=q^{\textrm{approx}}_{0l}(r) +\beta_l p_{0l}(r),
 \end{align}
 where $q^{\textrm{approx}}_{0l}(r)$ is an approximation that captures enough of the local behaviour of the full solution in the vicinity of the horizon to calculate the local contributions to either the scalar condensate or renormalised stress tensor. For the scalar condensate, sufficient local behaviour is captured through approximating $q_{0l}$ as:
 \begin{align}
   q^{\textrm{approx}}_{0l}(r)   =\frac{\Upsilon ^{1/4}}{[r^2 f(r)]^{1/4}}K_{0}(\rho_l\sqrt{\Upsilon }),
 \end{align}
 where $f(r)$ is the metric function (\ref{eq:metric}), $K_0(x)$ denotes the modified zeroth order Bessel function of the second kind and 
 \begin{equation}
   \Upsilon =\bigg(\int_{r_+}^{r} \frac{1}{\sqrt{r'^2 f(r')}} dr'\bigg)^2, \qquad \rho_l^2=\mu^2 r_+^2+\frac{1}{3} +l(l+1).
 \end{equation}

For the components of the RSET, which is constructed by taking derivatives of $W$, we must employ an approximation that captures more of the local behaviour of $q_{0l}$, such as:
\begin{equation}
{\widetilde {q}}^{\textrm{approx}}_{0l}(r) = 
\frac{\Gamma\left({\frac{1}{2} -\nu_l}\right)}{2^{3/2}[|\alpha| \Upsilon r^2 f(r)]^{1/4}}W_{\nu_l,0} \left(2\sqrt{|\alpha|} \Upsilon\right),
 \end{equation}
 where $W_{a,b}(x)$ denotes the Whittaker function of the second kind.
 
Since $p_{0l}(r)$ and $q_{0l}(r)$ satisfy the same second order linear differential equation (\ref{eq:Rad}), by integrating the Wronskian, we may also express $q_{0l}(r)$ in the form:
\begin{equation}
   q_{0l}(r)=p_{0l}(r)\int_{r}^{\infty} \frac{\mathrm{d}r'}{r'^2f(r') p_{0l}(r')^2}.
\end{equation}
Inserting this expression and the appropriate approximation into  (\ref{eq:approx}) and then expanding in the near-horizon limit leads to the following expressions for $\beta_{l}$ and $\widetilde {\beta}_l$:
\begin{align}
  \beta_{l}= & \int_{r_+}^{\infty} \frac{\mathrm{d}r'}{r'^2f(r')}\left(\frac{1}{p_{l0}(r')^2}-1\right)
  \nonumber \\ & +\frac{2}{r_+-r_{-}}\left[ \log(\rho _l) +\gamma\right], \nonumber \\ \nonumber
  \widetilde {\beta}_{l}= &\int_{r_+}^{\infty} \frac{\mathrm{d}r'}{r'^2f(r')}\left(\frac{1}{p_{l0}(r')^2}-1\right)\\ 
  &+\frac{1}{r_+-r_{-}}\left[ \log(8\sqrt{|\alpha|})+\psi\left(\frac{1}{2} +\frac{\rho_l}{8\sqrt{|\alpha|}}\right) +2\gamma\right] ,
\end{align}
where we have normalised the $p_{0l}(r)$ function to be unity on the horizon. For large $l$, the quantity $\beta_{l}$ is ${\mathcal {O}}(l^{-4})$ and $\widetilde {\beta}_l$ is ${\mathcal {O}}(l^{-6})$, hence the respective sums of these quantities converge.

\bibliography{charge}

\end{document}